%% file: main.tex
\documentclass[sigplan,nonacm]{acmart}
\author{Wenqi Jia*}
\affiliation{%
  \institution{UT Arlington}
  \country{}
}
\email{wenqi.jia@uta.edu}

\author{Zhewen Hu*}
\affiliation{%
  \institution{Texas A\&M University}
  \country{}
}
\email{zhewen@tamu.edu}

\author{Ying Huang*}
\affiliation{%
  \institution{UT Arlington}
  \country{}
}
\email{ying.huang@uta.edu}

\author{Jian Xu}
\affiliation{%
  \institution{UT Arlington}
  \country{}
}
\email{jian.xu@uta.edu}

\author{Siyi Wu}
\affiliation{%
  \institution{UT Arlington}
  \country{}
}
\email{siyi.wu@uta.edu}

\author{Yangming Zhang}
\affiliation{%
  \institution{UT Arlington}
  \country{}
}
\email{yangming.zhang@uta.edu}

\author{Yu Gong}
\affiliation{%
  \institution{Independent Researcher}
  \country{}
}
\email{yu3.gong@gmail.com}

\author{Stavros Kalafatis}
\affiliation{%
  \institution{Texas A\&M University}
  \country{}
}
\email{skalafatis-tamu@tamu.edu}

\author{Yuke Wang}
\affiliation{%
  \institution{Rice University}
  \country{}
}
\email{yuke.wang@rice.edu}

\author{Wei Niu}
\affiliation{%
  \institution{University of Georgia}
  \country{}
}
\email{wniu@uga.edu}

\author{Chengming Zhang}
\affiliation{%
  \institution{University of Houston}
  \country{}
}
\email{czhang59@central.uh.edu}

\author{Ang Li}
\affiliation{%
  \institution{University of Washington}
  \country{}
}
\email{uuudown@gmail.com}

\author{Sheng Di}
\affiliation{%
  \institution{Argonne National Labs}
  \country{}
}
\email{sdi1@anl.gov}

\author{Yuede Ji}
\affiliation{%
  \institution{UT Arlington}
  \country{}
}
\email{yuede.ji@uta.edu}

\author{Bo Fang}
\affiliation{%
  \institution{UT Arlington}
  \country{}
}
\email{bo.fang@uta.edu}

\author{Miao Yin$^\dagger$}
\affiliation{%
  \institution{UT Arlington}
  \country{}
}
\email{miao.yin@uta.edu}

\renewcommand{\shortauthors}{Jia and Yin, et al.}

\thanks{* Equal contribution.}
\thanks{$^\dagger$ Corresponding author.}
\input{preamble}
\newcommand{\abbr}{\textsf{Splaxel}}
\begin{document}
\title{\texorpdfstring{\abbr: Efficient Distributed Training of 3D Gaussian \underline{Spla}tting for Large-scale Scene Reconstruction via Pi\underline{xel}-level Communication}{\abbr: Efficient Distributed Training of 3D Gaussian Splatting for Large-scale Scene Reconstruction via Pixel-level Communication}}
\input{sec/0_abstract.tex}
\maketitle
\input{sec/1_introduction.tex}

\input{sec/2_background.tex}

\input{sec/3_motivation.tex}

\input{sec/4_design.tex}

\input{sec/5_evaluation.tex}
\input{sec/6_related_work.tex}
\input{sec/7_conclusion.tex}

\clearpage
\bibliographystyle{ACM-Reference-Format}
\bibliography{refs}
\clearpage
\input{sec/8_appendix.tex}

\end{document}

%% file: preamble.tex
\usepackage{tikz}
\newcommand*\Circled[2][gray!40]{%
	\tikz[baseline=(char.base)]{\node[
        shape=circle, draw=none,  thick, 
        fill=#1 ,inner sep=0.9pt] (char) 
    {\textcolor{black}{#2}}; 
}}

\usepackage{soul}
\usepackage{bm}
\usepackage{amsmath}
\usepackage[ruled,vlined,linesnumbered]{algorithm2e}
\usepackage{graphicx}
\usepackage{textcomp}
\usepackage{xcolor}
\usepackage{algorithm}
\usepackage{listings}
\usepackage{minted}
\usepackage{booktabs}
\usepackage{wrapfig}
\usepackage{subcaption}
\usepackage{multirow}
\usepackage{siunitx}
\usepackage{pbalance}
\usepackage[table]{xcolor}
\usepackage{makecell}

\def\BibTeX{{\rm B\kern-.05em{\sc i\kern-.025em b}\kern-.08em
    T\kern-.1667em\lower.7ex\hbox{E}\kern-.125emX}}

\setminted{
  linenos,
  frame=none,                 %
  numbersep=6pt,
  xleftmargin=1.0em,          %
  baselinestretch=1.0,
  autogobble=true,
  breaklines,
  breaksymbolleft=,           %
  breaksymbolsepleft=0pt,
  numberblanklines=false,
  stripnl=true
}

%% file: sec/0_abstract.tex
\begin{abstract}
3D Gaussian Splatting (3DGS) enables high-fidelity and real-time 3D scene reconstruction, but scaling training to large-scale scenes requires optimizing hundreds of millions of Gaussians across multiple GPUs. Existing distributed approaches either partition scenes into isolated regions, causing global inconsistency, or rely on global Gaussian-level exchanges, which lead to substantial growth in inter-GPU communication and quickly dominate iteration time.

We propose \textsf{Splaxel}, a communication-efficient distributed 3DGS training framework based on pixel-level local rendering and global composition. Instead of synchronizing Gaussians, each GPU renders its local subset and exchanges only partial pixel values, maintaining mathematical consistency while keeping communication cost stable as the scene size increases. \textsf{Splaxel} further reduces pixel-level redundancy through geometric and transmittance visibility prediction and improves GPU utilization via conflict-free camera-view consolidation. Evaluated on large-scale datasets with up to 120M Gaussians, \textsf{Splaxel} achieves up to 7.6$\times$ speedup over the state-of-the-art distributed 3DGS framework while preserving high reconstruction quality. 
\end{abstract}

%% file: sec/1_introduction.tex
\section{Introduction}

\begin{figure}[t]
    \centering
    \includegraphics[width=0.95\linewidth]{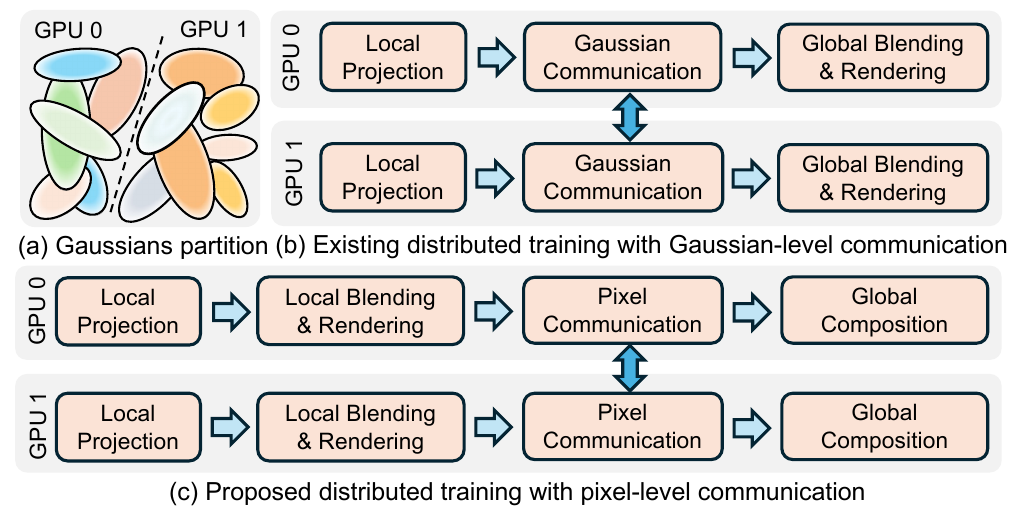}
    \vspace{-5mm}
    \caption{Conceptual comparison between pixel-level communication and Gaussian-level communication schemes.}
    \Description{A figure showing a conceptual comparison between pixel-level communication and Gaussian-level communication schemes.}
    \label{fig:intro-conceptual-diff}
    \vspace{-3mm}
\end{figure}

3D scene reconstruction supports a wide range of applications, including autonomous driving~\cite{li2019aads,ost2021neural, zielonka2025drivable, zhou2024drivinggaussian}, digital twins~\cite{deng2021systematic}, and augmented reality~\cite{franke2025vr, jiang2024vr}, where modeling fine geometric and appearance detail is critical~\cite{kulhanek2024wildgaussians}.
To facilitate such a process, 3D Gaussian Splatting (3DGS)~\cite{kerbl20233d, lu2024scaffold, ren2024octree} has emerged as the dominant solution for high-fidelity, real-time scene rendering. 3DGS trains by projecting each 3D Gaussian, defined by its position, covariance, color, and opacity, onto the target camera plane, where they are depth-sorted and splat-rendered with alpha blending to form the image. The rendering loss is then backpropagated through the splatting pipeline to update all Gaussian parameters~\cite{kerbl20233d, li20253d}. 
By optimizing millions of those explicit 3D Gaussians, 3DGS captures rich spatial details and achieves significantly higher rendering efficiency compared to conventional neural implicit representations~\cite{mildenhall2021nerf, lee2023neurex, li2022rt, song2023cambricon}.

\begin{figure*}[t!]
    \centering
    \includegraphics[width=\linewidth]{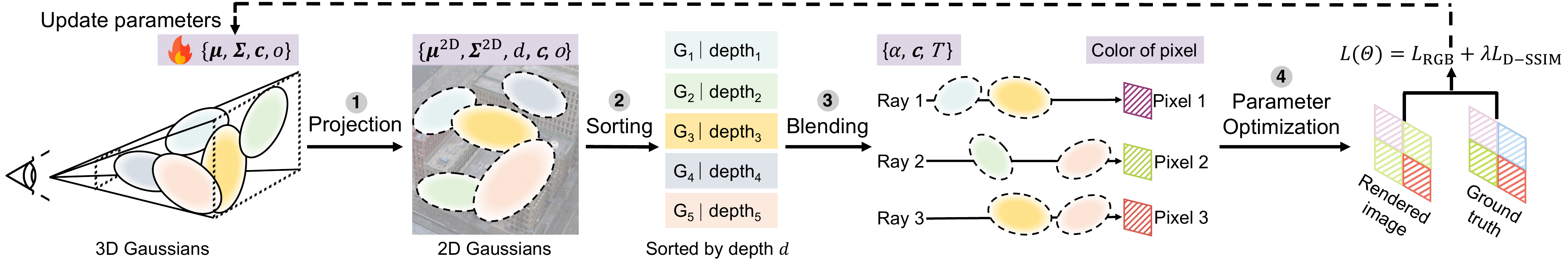}
    \vspace{-8.5mm}
    \caption{The standard training pipeline of 3D Gaussian Splatting.}
    \Description{The standard training pipeline of 3D Gaussian Splatting.}
    \label{fig:background}
\end{figure*}

Despite its rendering efficiency, scaling 3DGS to large scenes is fundamentally challenging due to its dynamic, view-dependent access patterns and the resulting inefficiencies in distributed training~\cite{zhang2024lp, zhou2024feature, zhang2025gaussianspa, yan2024gs}. 
To scale up 3DGS for large-scale 3D scene reconstruction~\cite{knapitsch2017tanks, jiang2025horizon, chen2024gigags}, the primary solution is to leverage distributed training to break the memory barrier of optimizing massive numbers of Gaussians. However, 3DGS exhibits a dynamic and imbalanced access pattern that depends on the current random camera view, where it activates distinct subsets of Gaussians in different views~\cite{clm2026asplos, papantonakis2024reducing}. Moreover, the increased number of Gaussians in large-scale scenes significantly intensifies this pattern. 
Consequently, existing distributed training frameworks for deep neural networks (DNNs), e.g., DeepSpeed~\cite{rasley2020deepspeed}, PyTorch DDP~\cite{li2020pytorch}, Alpha~\cite{zheng2022alpa}, which assume static and balanced tensor computations, are incompatible with 3DGS.

To tackle these challenges, existing efforts can be broadly categorized into two main directions. \ul{\textit{i) Algorithm optimization.}} Existing works~\cite{lin2024vastgaussian, liu2024citygaussian, chen2024dogs, kerbl2024hierarchical} in this category leverage divide-and-conquer strategies, partitioning the scene into independent regions, fully optimizing each region separately across GPUs, and finally merging them for complete reconstruction.
While these works can optimize Gaussian parameters locally with the dynamic camera views, they sacrifice global consistency. Consequently, it leads to visible discontinuities across the reconstructed scene as each partition optimizes its Gaussians in isolation without considering neighboring regions.
\ul{\textit{ii) System optimization.}} In this direction, Grendel~\cite{zhao2025scaling3dgs} is the only work that globally preserves the 3DGS pipeline and consistency.
Grendel randomly distributes Gaussians across devices, where each device determines the required Gaussians based on the current camera view and requests the missing Gaussians from other devices through all-to-all communication. While this design maintains global consistency, it faces significant scalability challenges in large-scale scenes. As the number of Gaussians grows, such a global \textit{Gaussian-level} exchange becomes increasingly expensive, leading to a substantial increase in the communication ratio for inter-GPU data transfer. Consequently, communication dominates training iteration time and severely limits 3DGS's scalability and throughput.

To address the above issues, we consider the opportunity of a \textit{pixel-level} communication scheme, where each GPU renders locally based on its local Gaussian partition and exchanges only the rendered partial pixels for global blending, as shown in Fig.~\ref{fig:intro-conceptual-diff}. 
Pixel-level communication enables constant data transmission volume across GPUs, regardless of the number of Gaussians involved. 
Such property alleviates the burden of substantial Gaussians in large-scale scenes. Despite the potential to mitigate communication bottleneck, three main challenges exist. 
i) Global inconsistency. At the pixel level, local rendering lacks the global Gaussian information, which may lead to a Gaussian disorder in the blending phase. 
ii) Pixel-level redundancy. A naive pixel-level communication scheme leads to spatial redundancy and saturation redundancy due to geometric invisibility and transmittance saturation, respectively, resulting in a majority of transmissions being zeros, which implies significant bandwidth waste on data that never affects the rendering result. 
iii) Insufficient GPU utilization. Most camera views only cross a subset of the Gaussian partitions, but all GPUs must participate in the computation for a single view under the naive scheduling strategy. This leads to a high GPU idle rate.

In this paper, we propose \text{\abbr}, enabling efficient distributed 3DGS training for large-scale scene reconstruction via pixel-level communication. \text{\abbr} leverages convex-based partition to ensure the global consistency. It reduces spatial redundancy by projecting the vertices of overlapping 3D space between the view frustum and the convex Gaussian partition onto the rendering plane without extra communication. Moreover, \text{\abbr} enhances GPU utilization through an efficient conflict-free camera view and GPU scheduling mechanism. Our contributions are summarized as follows:
\begin{itemize}
    \item We analyze the bottleneck ($\S$\ref{sec:bottleneck}) in the existing distributed 3DGS training framework with Gaussian-level communication, identifying the opportunities and challenges of a pixel-level communication scheme.
    
    \item We propose a principled 3DGS distributed training framework based on pixel-level communication ($\S$\ref{sec:design-pixel-level}), with a reformulation of the Gaussian rendering and parameters update paradigm to maintain mathematical consistency. 
    
    \item We present two novel techniques ($\S$\ref{sec:design-redundancy-reduction}, $\S$\ref{sec:design-scheduling}) that reduce pixel-level redundancy by predicting geometric and transmittance invisibility, and improve GPU utilization with conflict-free camera view consolidation.

    \item We conduct extensive evaluations ($\S$\ref{sec:eval}) on multiple large-scale datasets, including the largest publicly available ones, Big City Street and Aerial, with 120M Gaussians, on which \text{\abbr} outperforms the existing SOTA by 7.6$\times$ and 2.9$\times$ speedups with 8 GPUs, respectively.
    
\end{itemize}

%% file: sec/2_background.tex
\section{Background}
\label{sec:background}

\begin{figure}[t]
    \centering
    \includegraphics[width=1.0\linewidth]{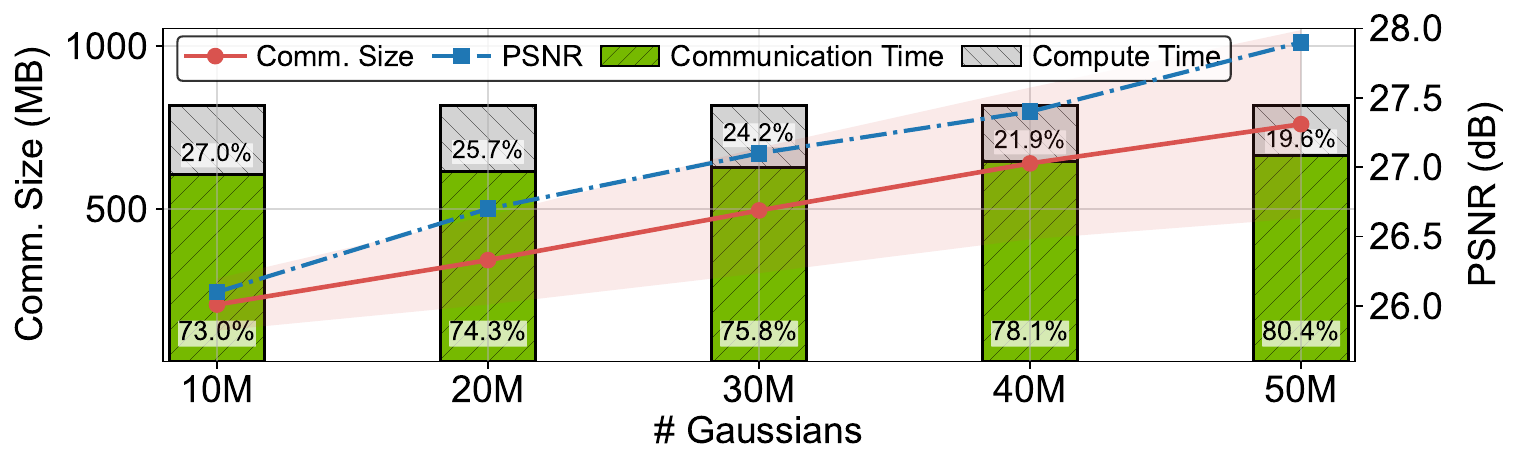}
    \vspace{-9mm}
    \caption{Communication cost and time breakdown per iteration with 8 GPUs and PSNR w.r.t. the number of Gaussians. The red shaded region denotes the transmitted mean $\pm$ 1$\sigma$. }
    \Description{A plot showing communication cost per iteration on 8 GPUs and PSNR as the number of Gaussians increases.}
    \label{fig:motivaton-comm-size-scale}
    \vspace{-3mm}
\end{figure}

\begin{figure}[t]
    \centering
    \includegraphics[width=\linewidth]{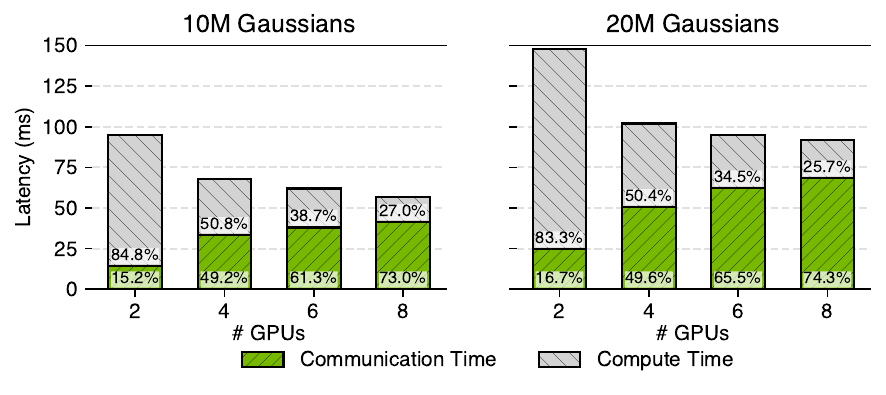}
    \vspace{-9.5mm}
    \caption{Communication bottleneck. Communication rises from 15\% to over 70\% per iteration as scene and GPU scale.}
    \Description{A plot illustrating the communication bottleneck.}
    \label{fig:motivation-comm-breakdown}
\end{figure}

3D Gaussian Splatting (3DGS) \cite{kerbl20233d, yang2024deformable, charatan2024pixelsplat}  encodes the target scene with a collection of 3D Gaussians. Each Gaussian is represented by its position $\boldsymbol{\mu} \in \mathbb{R}^3$ for 3D position, covariance matrix $\bm{\Sigma}\in \mathbb{R}^{3 \times 3}$,  as well as the opacity value $o$ and the color $\bm{c}$. The covariance matrix $\bm{\Sigma}$ is generally composed of $\bm{\Sigma} = \bm{R} \bm{S} \bm{S}^\top \bm{R}^\top$ with a rotation matrix $\bm{R} \in \mathbb{R}^{3 \times 3}$ and a scaling factor $\bm{S} \in \mathbb{R}^{3 \times 3}$, In summary, each Gaussian is parametrized by $\bm{\Theta}_i = \{\bm{\mu}_i,\bm{R}_i, \bm{S}_i, o_i, \bm{c}_i\}$.

Fig. \ref{fig:background} illustrates the training process to optimize each Gaussian's parameters through the following four main steps:

\textit{\Circled{1} Projection.} 3D Gaussians outside the camera’s field of view are first discarded through frustum culling according to its mean position $\bm{\mu}$. Each visible 3D Gaussian is then projected into an elliptical 2D Gaussian on the view plane, resulting in 2D attributes, such as depth $d$, 2D position $\bm{\mu}^{\text{2D}}$ and 2D covariance $\bm{\Sigma}^{\text{2D}}$. The projected 2D covariance matrix is computed as
$\bm{\Sigma}^{\text{2D}} = \bm{J} \bm{W} \bm{\Sigma} \bm{W}^{\top} \bm{J}^{\top},$
where $\bm{W}$ represents the view transformation matrix, and $\bm{J}$ is the Jacobian of the affine approximation
of the projective transformation. Typically, a pixel will be covered by multiple Gaussians, forming a one-to-many relationship between the pixel and the Gaussians.

\textit{\Circled{2} Sorting.} To produce physically correct rendering results that follow light propagation mechanism, 3D Gaussians are sorted based on their distance from the camera, i.e., depth $d$.

\textit{\Circled{3} Blending \& Rendering.} 3DGS leverage $\alpha$-blending to compute the pixel color of the rendered image. For the $i$-th Gaussian, the rendering opacity $\alpha_i$ is calculated by\vskip -4mm
\begin{equation}
\alpha_i = o_i \exp(
  -\tfrac{1}{2} (\bm{x} - \bm{\mu}_i^{\text{2D}})^{\mathrm{T}}
  {(\bm{\Sigma}_i^{\text{2D}})}^{-1}
  (\bm{x} - \bm{\mu}_i^{\text{2D}})
).
\end{equation}
Then, the pixel color is rendered by accumulating the Gaussians' color as
\begin{equation}
    C_p = \sum_{i} \bm{c}_i \, \alpha_i T_i,
    \label{eq:color_rendering}
\end{equation}
and $T_i = \prod_{j=1}^{i-1} (1 - \alpha_j)$ is the transmittance of $i$-th Gaussian.

\textit{\Circled{4} Parameter Optimization.} After rendering the  image, the loss between the input image and the rendered image in the pixel space is calculated by
$
L = L_\text{RGB} + \lambda L_\text{D-SSIM},
$
where $L_\text{RGB}$ denotes the RGB rendering error, and $L_\text{D-SSIM}$ denotes the structural dissimilarity derived from the Structural Similarity Index Measure (SSIM).

%% file: sec/3_motivation.tex
\section{Motivation}
\label{sec:motivation}

In this section, we introduce the fundamental bottleneck that needs to be addressed for large-scale 3D reconstruction and explain our hypothesis for mitigating this bottleneck. 

\subsection{Observation: Communication Bottleneck in Large-scale Scene Reconstruction}
\label{sec:bottleneck}

Our observation reveals a communication bottleneck inherent in current distributed 3DGS training frameworks. Both scene scaling (increasing the number of Gaussians) and system scaling (increasing the number of GPUs) lead to unexpected, substantial growth in communication during inter-GPU data transfer. As a result, communication dominates iteration time and severely limits scalability and throughput.

\begin{figure}[t]
    \centering
    \includegraphics[width=0.95\linewidth]{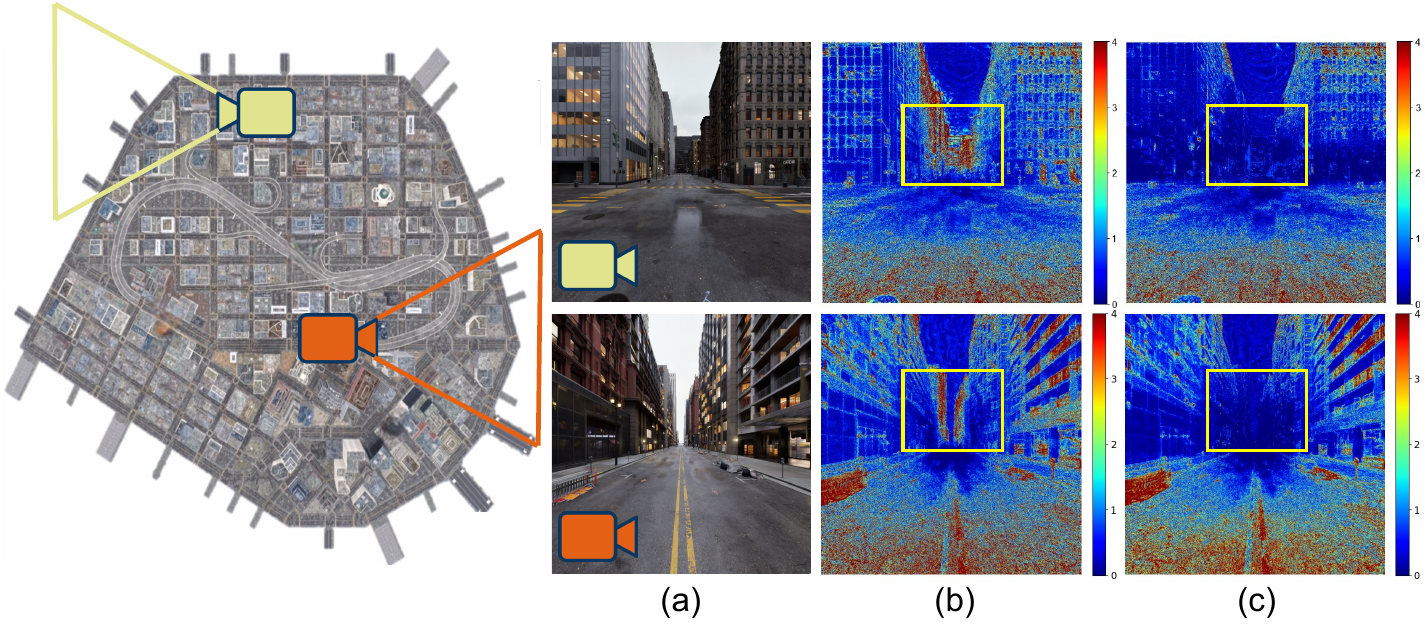}
    \vspace{-5mm}
    \caption{Rendering error maps at two example locations of the Small City Street dataset. (a) Ground truth images. (b) Training with opacity-based Gaussian filtering. (c) Regular training with unbounded Gaussians.} 
    \Description{Rendering error maps at two example locations in the Small City Street dataset.}
    \label{fig:motivation-naive-solution}
\end{figure}

Achieving high-quality reconstruction for large-scale scenes requires optimizing tens to hundreds of millions of Gaussians, posing significant challenges for distributed 3DGS training~\cite{ye2025gs, chen2025dashgaussian, fan2025momentum}. To understand the bottleneck in the state-of-the-art framework, Grendel~\cite{zhao2025scaling3dgs}, we conduct a breakdown analysis on an 8-GPU platform with NVIDIA RTX 6000 Ada GPUs using the large-scale MatrixCity dataset~\cite{li2023matrixcity}. Fig.~\ref{fig:motivaton-comm-size-scale} shows that as the number of Gaussians increases from 10M to 50M, per-iteration communication grows from 200 MB to 800 MB and dominates runtime, accounting for 73.0\%-80.4\%. This trend indicates that larger scenes significantly increase the number of visible Gaussians exchanged across GPUs. Fig.~\ref{fig:motivation-comm-breakdown} further shows that communication overhead worsens with more GPUs. Although local computation decreases with parallelism, cross-GPU synchronization grows much faster and quickly dominates iteration time. For example, with 20M Gaussians, increasing GPUs from 2 to 8 raises per-GPU communication time from 24 ms to 68 ms, while the communication ratio increases from 16.7\% to 74.3\%.

These results reveal a fundamental communication bottleneck in existing distributed 3DGS frameworks: both scene scaling and system scaling lead to substantial increases in inter-GPU data transfer, causing communication to dominate runtime and limiting scalability.

\begin{figure}[t]
\vspace{1mm}
    \centering
    \includegraphics[width=0.9\linewidth]{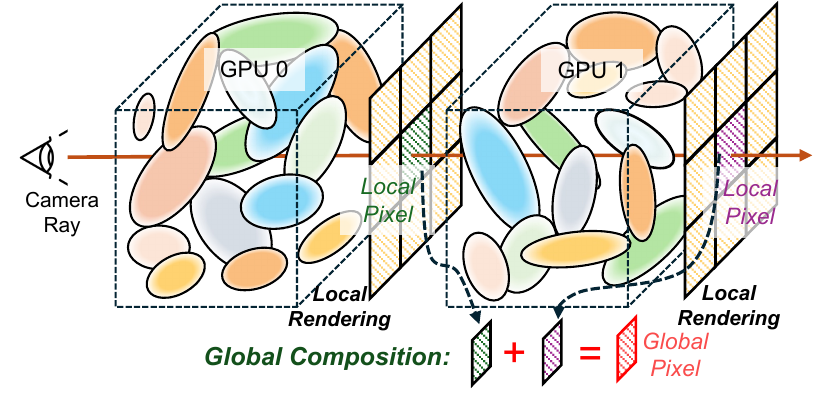}
    \vspace{-5mm}
    \caption{Each GPU accumulates the color and transmittance from its local Gaussians into per-pixel aggregates during local rendering. The local rendered pixel encapsulates all necessary information for cross-GPU composition, eliminating the need to transmit individual Gaussians and maintaining a constant communication cost regardless of scene complexity.}
    \Description{An illustration of pixel-level aggregation for cross-GPU composition.}
    \label{fig:motivation-pixel-level}
    \vspace{-3mm}
\end{figure}

\subsection{Root Cause of Communication Bottleneck: View-dependent Gaussian Access}

\setlength{\columnsep}{8pt}
\begin{wrapfigure}{r}{0.39\linewidth}
    \centering
    \vspace{-4mm}
    \includegraphics[width=\linewidth]{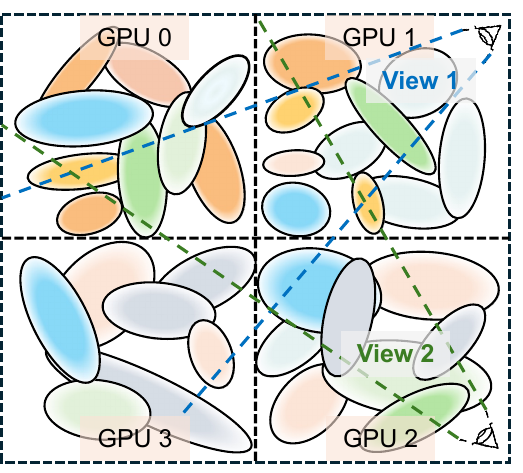}
    \vspace{-8mm}
    \caption{The requested Gaussians are dynamically dependent on the current views.}
    \Description{An illustration showing that the set of requested Gaussians changes dynamically with the current view.}
    \vspace{-3mm}
    \label{fig:motivation-view-dependent}
\end{wrapfigure}
In the existing distributed training paradigm, for each target view in the dataset, each GPU must request the required projected Gaussians that are not on itself to render the whole image to calculate the gradient and update the Gaussian parameters.
In large scenes containing tens of millions of Gaussians, a single target view may involve millions of cross-GPU requests. 
Consequently, such a Gaussian-level communication scheme causes significantly increased communication overhead. 
Additionally, the number of cross-view GPUs increases in proportion to the number of GPUs. When more GPUs participate in training, the Gaussians required for a given view are scattered across a larger set of devices. Each GPU must therefore retrieve remote Gaussians from more peers, multiplying the total amount of data exchanged. This effect creates a nonlinear growth in communication cost with respect to both the number of Gaussians and the number of GPUs. Due to the nature that per-iteration Gaussians are dependent on the dynamic views, which are randomly distributed from all angles in the large-scale datasets, it is unable to reduce the amount of Gaussian communication for all views via an optimized partition as demonstrated in Fig. \ref{fig:motivation-view-dependent}.
Furthermore, the communication is tightly synchronized with the rendering and backpropagation pipeline. Each GPU must wait for the arrival of remote Gaussian data before proceeding with global pixel composition, creating serialization points that further amplify iteration delay. The overall system throughput, therefore, becomes bounded by inter-GPU communication.

\subsection{Can Reducing Gaussians Mitigate Bottleneck?}

A natural approach to alleviate the communication bottleneck is to reduce the number of transmitted Gaussians, but this introduces a fundamental trade-off between efficiency and rendering correctness. Specifically, each GPU transmits only the Gaussians whose opacity exceeds a predefined threshold, while skipping those with low transparency values that contribute less to the rendered image. This simple rule effectively discards Gaussians from inter-GPU communication, significantly reducing both the data volume and transmission time per iteration. However, this approach compromises global rendering consistency. The opacity of a Gaussian is highly view-dependent and context-sensitive: Gaussians that appear nearly transparent in one view may still play a critical role in another. By discarding low-opacity Gaussians globally, the framework introduces inconsistencies across views, as different GPUs may operate with incomplete or mismatched Gaussian sets, leading to incorrect pixel composition and visible rendering artifacts. 
As shown in Fig. \ref{fig:motivation-naive-solution}, on the Small City Street dataset, when transmitting only Gaussians whose opacity exceeds a certain threshold, the rendering error is significantly amplified compared to regular distributed training. This trade-off between communication efficiency and reconstruction fidelity highlights the fundamental limitation of Grendel~\cite{zhao2025scaling3dgs}, motivating a more principled distributed training framework for large-scale reconstruction.

\begin{figure}[t]
    \centering
    \vspace{-1mm}
    \includegraphics[width=1.0\linewidth]{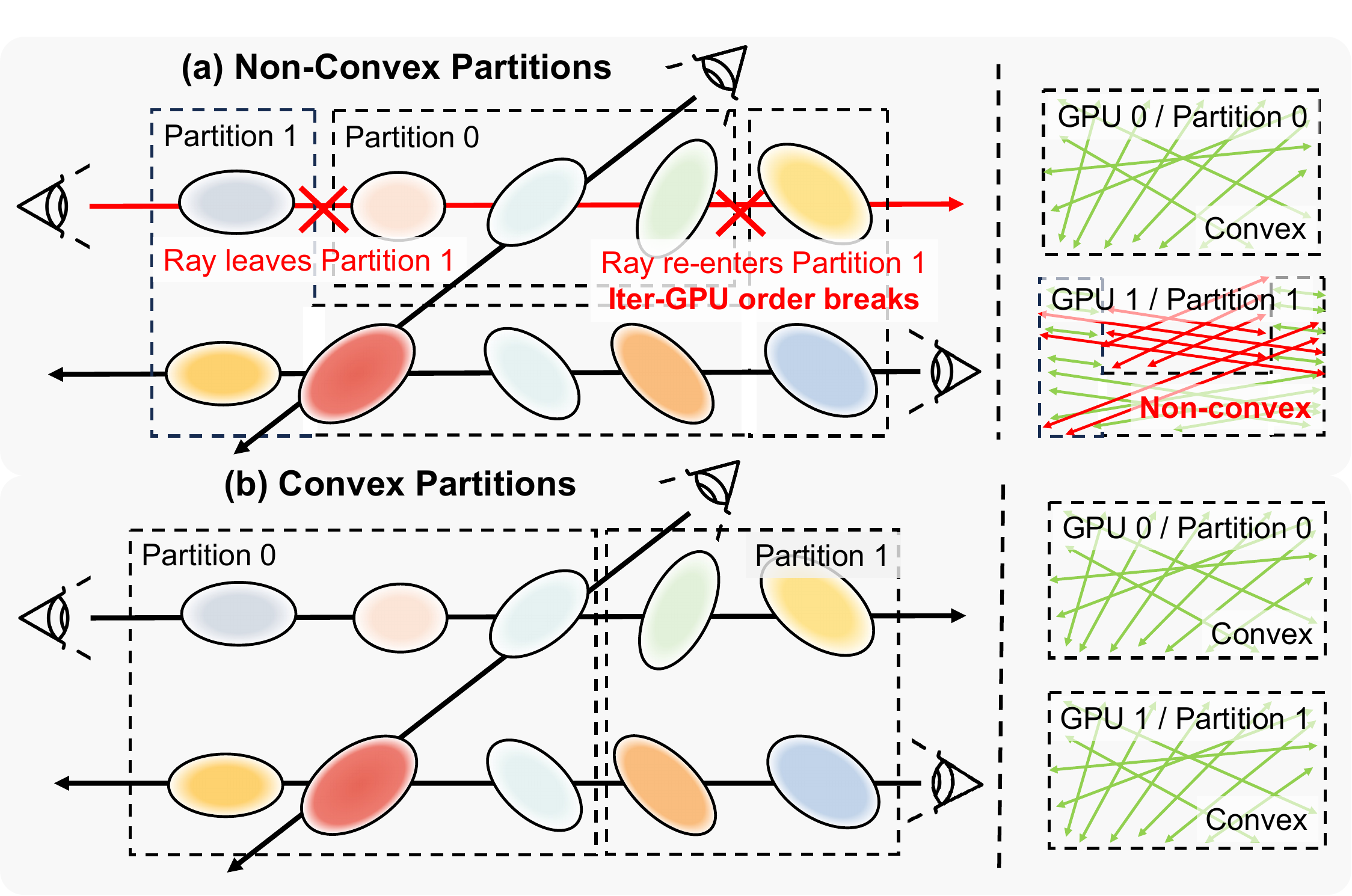}
    \vspace{-7mm}
    \caption{(a) Local rendering in non-convex partitions cannot guarantee the global blending order. (b) Convex partitions ensure consistent global inter-GPU depth ordering.}
    \Description{A comparison showing that convex partitions preserve consistent global depth ordering, while non-convex partitions do not.}
    \label{fig:design-convex-partition}
    \vspace{-1mm}
\end{figure}

\subsection{Opportunity: Pixel-level Communication}

To fundamentally mitigate the bottleneck, we consider the opportunity, \textit{can we transmit local rendering intermediates instead of the Gaussians?} Specifically, each GPU independently accumulates the contributions of its local Gaussians and performs rendering to obtain an intermediate image. Then, all the corresponding pixels from intermediate images are aggregated along the same camera ray to render the final image. As illustrated in Fig.~\ref{fig:motivation-pixel-level}, GPU~0 and GPU~1 compute their local rendering results independently, and GPU~1 only needs the ``local pixel" (red) from GPU~0 to render the pixel at the same ``local pixel" (blue) on the camera ray. In this way, the communication is no longer relevant to the number of Gaussians. Instead, the communication cost of such a pixel-level communication scheme is approximately constant with respect to the resolution, thereby eliminating the tremendous Gaussian transmission in large-scale scenes.

%% file: sec/4_design.tex
\section{Pixel-level Communication for Efficient Distributed 3DGS Training}
\label{sec:design}

\subsection{Overview}

Based on the analysis and opportunity presented in Section \ref{sec:motivation}, we propose \text{\abbr}, an efficient distributed 3DGS training framework for high-quality large-scale scene reconstruction via pixel-level communication. The key innovation in the proposed design is reformulating the distributed 3DGS training to expose pixel-level communication, which inherently targets large-scale scene reconstruction tasks. 
This reformulation, discussed in Section \ref{sec:design-pixel-level}, provides a principled foundation for pixel-level distributed 3DGS training that differs fundamentally from the prior Gaussian-level approach.

\begin{figure}[t]
\centering
\includegraphics[width=1.0\linewidth]{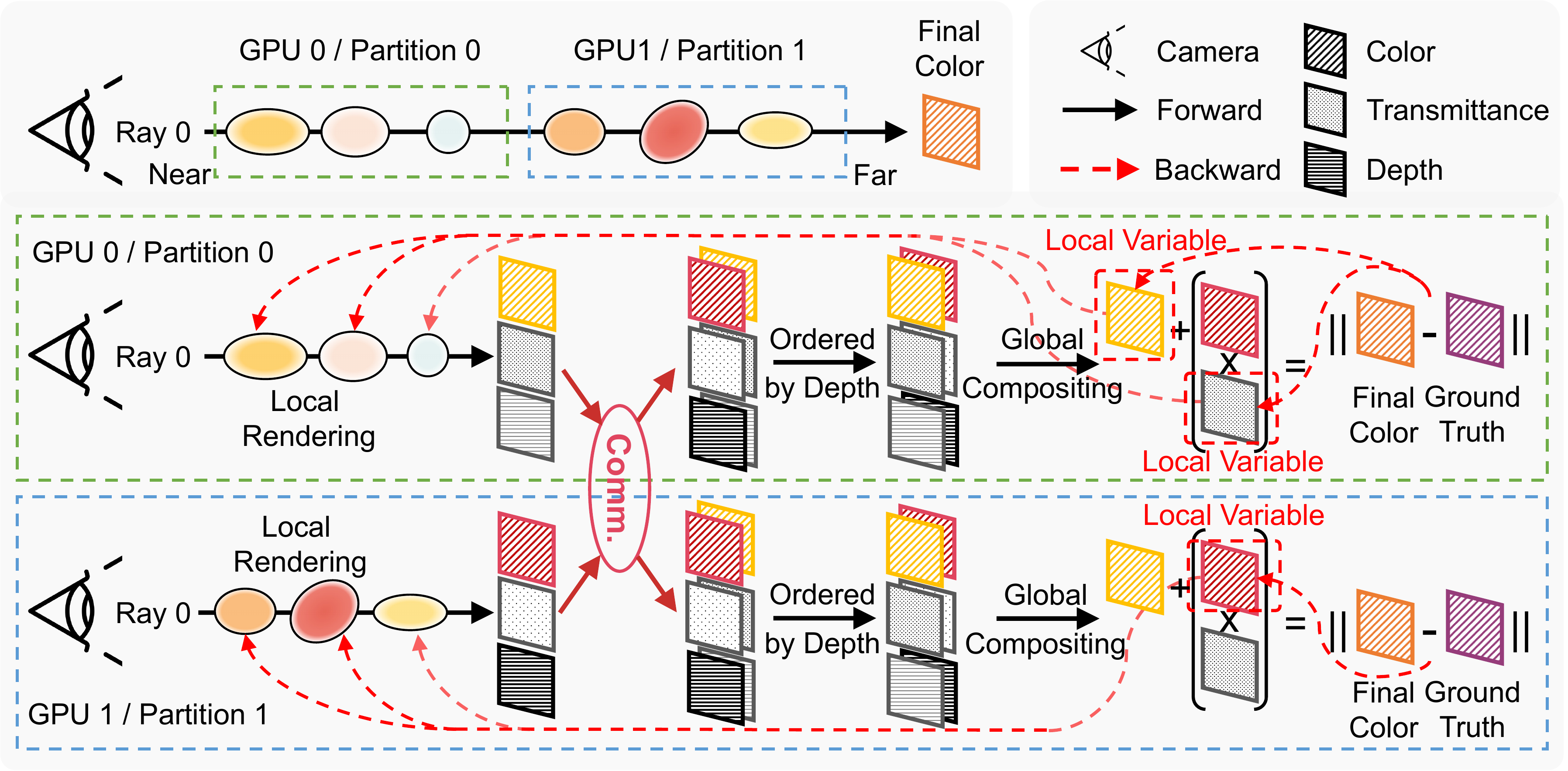}
\vspace{-8mm}
\caption{Distributed rendering with pixel-level communication.
Each GPU exchanges per-pixel color, transmittance, and depth; 
The received pixels are composited in global depth order for consistent rendering, and gradients propagate locally without further communication. }
\Description{An illustration of distributed rendering with pixel-level communication and global depth-ordered composition across GPUs.}
\label{fig:design-order-and-blend}
\vspace{-1.5mm}
\end{figure}

Adopting a pixel-level communication scheme introduces challenges of redundant inter-GPU pixel transmission and insufficient GPU utilization. We develop two novel techniques to address those challenges and fully unlock the potential of the proposed pixel-level communication scheme. Section \ref{sec:design-redundancy-reduction} presents a communication-free approach that predicts the invisible pixels before and during training, respectively, thereby offering a significant reduction in spatial and saturation redundancy. 
Section \ref{sec:design-scheduling} presents an efficient scheduling mechanism with conflict-free view consolidation, which leverages the disjoint nature of workload distribution to enable concurrent execution of multiple views. This mechanism removes GPU bubbles and improves overall throughput by reducing synchronization across different camera views.

\subsection{Pixel-Level Communication Scheme}
\label{sec:design-pixel-level}

\noindent\ul{\textbf{Challenge: How to Maintain Inter-GPU Blending Order?}}
To employ pixel-level communication for 3DGS distributed training, the key challenge is to maintain the global blending order of Gaussians. Recalling the blending step in Section \ref{sec:background}, the color of each rendered pixel is computed through the sequential accumulation of colors from \textit{depth-ordered} Gaussians along the camera ray, where the transmittance at each step depends on all previously accumulated opacity. For pixel-level communication, each GPU has no access to the Gaussians residing on other GPUs. The local pixels (partial to the final pixels) generated by a GPU need to be consistent with the global blending order for the training algorithm to yield the correct rendering results. As illustrated in Fig. \ref{fig:design-convex-partition}a, the red camera ray enters Partition 1 twice, and the inter-GPU depth order is broken when aggregating local pixels from GPU 0 and GPU 1. To maintain the global order, GPU 1 must render two local images based on the Gaussians on the left and right sides of Partition 0, respectively, resulting in the doubled communication cost for Partition 1.

\noindent\ul{\textbf{Opportunity: Convex Partitioning Enables Globally Ordered Local Rendering.}}
To overcome the challenge, we identify an opportunity to preserve global blending order by exploiting geometric properties. We find that \textit{if all the partitions are convex, the global blending order is guaranteed.} A convex partition is defined as a region in which any line segment connecting two points within it remains entirely inside. When the scene is divided into convex regions, each pixel’s ray intersects with any GPU’s region at most once, resulting in a globally ordered accumulation of Gaussian contributions. As shown in Fig.~\ref{fig:design-convex-partition}b, all the camera rays enter and exit both Partition 0 and 1 only once. Therefore, GPU 0 and 1 can perform local blending based on local Gaussians by depth only once, and the global aggregation of per-GPU pixels along the ray preserves the global blending ordering without extra communication. 
\text{\abbr} employs the axis-aligned bounding box (AABB) method to obtain convex partitions.

\begin{figure*}[t]
\centering
\vspace{-1.5mm}
\includegraphics[width=\linewidth]{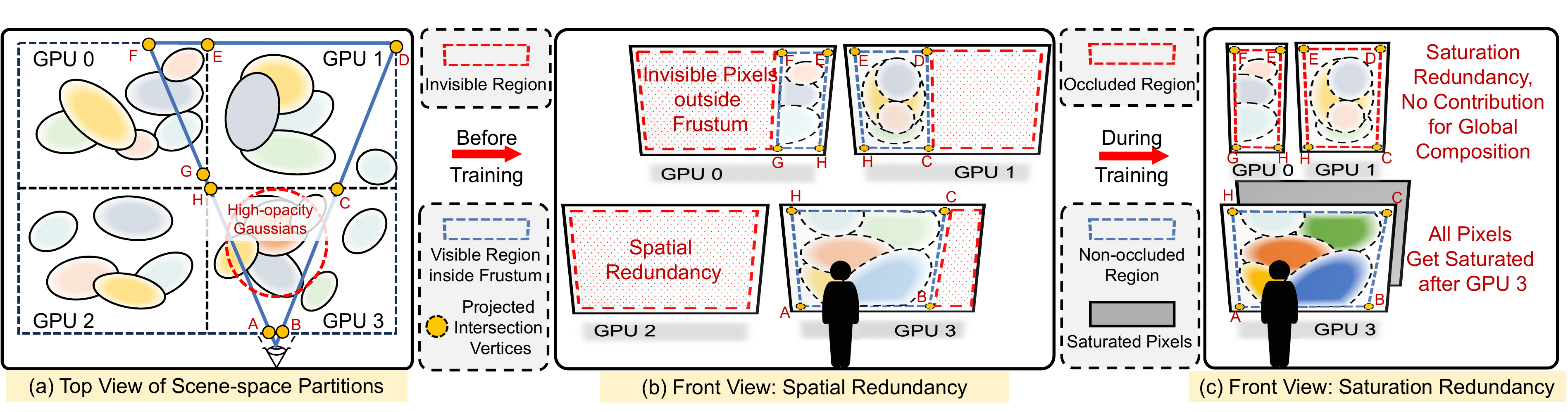}
\vspace{-8.5mm}
\caption{Spatial and saturation redundancy.
(a) Each GPU covers only a portion of the scene geometry, leaving invisible regions outside its frustum that cause \textit{spatial redundancy}.
(b) These invisible regions lead to redundant local pixel rendering and communication, as shown in the front view.
(c) During training, some Gaussians become highly opaque, producing \textit{saturation redundancy} where later GPUs no longer contribute to the global composition. }
\Description{An illustration of spatial redundancy from invisible regions and saturation redundancy from highly opaque Gaussians during distributed rendering.}
\label{fig:design-spatial-redundancy}
\vspace{-1mm}
\end{figure*}

\noindent\ul{\textbf{Design of Distributed Training with Pixel-level Communication.}}
We design the pixel-level communication-based distributed training framework as shown in Fig.~\ref{fig:design-order-and-blend}. The framework consists of three steps: \textit{local rendering}, \textit{global composition}, and \textit{backward propagation}. Among them, local rendering follows the standard 3DGS rendering procedure within each GPU's local partition, while our key contribution is to enable mathematically consistent cross-GPU training through pixel-level communication, including the global composition of partial pixels and the corresponding backward propagation.

\textit{\Circled{1} Local Rendering.} This step is mainly to compute each GPU’s partial pixel contributions by blending its locally stored Gaussians along each camera ray in depth order.
We define $\mathcal{Q}^m$ as the set of Gaussians on the $m$-th GPU, on which the local rendering process is performed within the corresponding convex partition. To render the partial pixel color $C_p^m$ for pixel $p$ on GPU $m$, the Gaussians associated with the camera ray of that pixel are first sorted locally by depth. Then, the color is rendered by the local blending as:\vskip -3.3mm
\begin{equation}
C^{m}_{p} = \sum_{i \in \mathcal{Q}^m} \bm{c}_i \, \alpha_i \prod_{j=1}^{i-1} (1 - \alpha_j).
\end{equation}
To aggregate all partial pixels from associated GPUs for the final rendering image, we also need to calculate the pixel-wise partial transmittance $T_p^m$ and the partial depth $D_p^{m}$ by \vskip -3.7mm
\begin{equation}
T_p^m = \prod_{i \in \mathcal{Q}^m} (1 - \alpha_i), D_p^m = \sum_{i \in \mathcal{Q}^m} d_i \prod_{j=1}^{i-1} (1 - \alpha_j). 
\end{equation}
Such information is transmitted to the GPUs that the camera ray crosses, along with the partial pixels for maintaining the global order.

\textit{\Circled{2} Global Composition.} This step composes partial pixels from multiple GPUs into the final image while preserving the global rendering order. For the pixel $p$, we first sort the colors according to their partial depth $D_p^m$, and then traverse the GPUs from front to back along with the camera ray to accumulate the local rendering color
with partial transmittance $T_p^m$. This will lead to a global rendering color $C_p$: \vskip -3mm
\begin{equation}
C_p = \sum_{m}^M C^m_p \prod_{\substack{ k < m}} T_p^k.
\label{eq:global_rendering}
\end{equation}
Eq.~\ref{eq:global_rendering} holds equivalently to Eq.\ref{eq:color_rendering} for all $p$. Fig.~\ref{fig:design-order-and-blend} shows only the local rendering color, partial depth, and partial transmittance need to be transmitted in the proposed training framework. Once the rendering resolution is set, the communication size is also fixed and independent to the number of Gaussians.

\textit{\Circled{3} Backward Propagation.} This step backpropagates gradients through the globally composed pixels to update each GPU's local Gaussians. After global rendering, we can calculate the loss $L$ between the globally rendered image and the ground truth, and compute the gradients to update the Gaussian parameters. In the pixel-level communication, since the color is obtained by combining partial pixels from multiple GPUs, we reformulate the gradient backpropagation accordingly. Similar to the forward pass, the per-Gaussian gradient backpropagates through the intermediate partial pixel on each GPU. Specifically, for all Gaussians on GPU $m$, i.e., $i\in\mathcal{Q}^m$, the gradient to $\alpha_i$ and $\bm{c}_i$, $\frac{\partial L}{\partial \alpha_i} = \frac{\partial L}{\partial C_p} \frac{\partial C_p}{\partial \alpha_i}$ and $\frac{\partial L}{\partial \bm{c_i}} = \frac{\partial L}{\partial C_p} \frac{\partial C_p}{\partial \bm{c_i}}$, can be calculated by 
\begin{align}
  \frac{\partial C_p}{\partial \alpha_i}
  &= \frac{\partial C_p^m}{\partial \alpha_i} \prod_{\substack{k < m}} T^k_p + \frac{\partial T^m}{\partial \alpha_i} \sum_{k > m} C^k_p  \prod_{\substack{ s < k, s \neq m}} T^s_p, \\
  \frac{\partial C_p}{\partial \bm{c_i}}
  &= \frac{\partial C_p^m}{\partial \bm{c}_i} \prod_{\substack{k < m}} T^k_p. 
\end{align}
As each GPU retains the local rendering color and partial transmittance of other GPUs, the parameters can be optimized without any additional cross-GPU communication.
\begin{figure*}[t]
\centering
\begin{subfigure}[t]{0.48\linewidth}
    \centering
    \includegraphics[width=\linewidth]{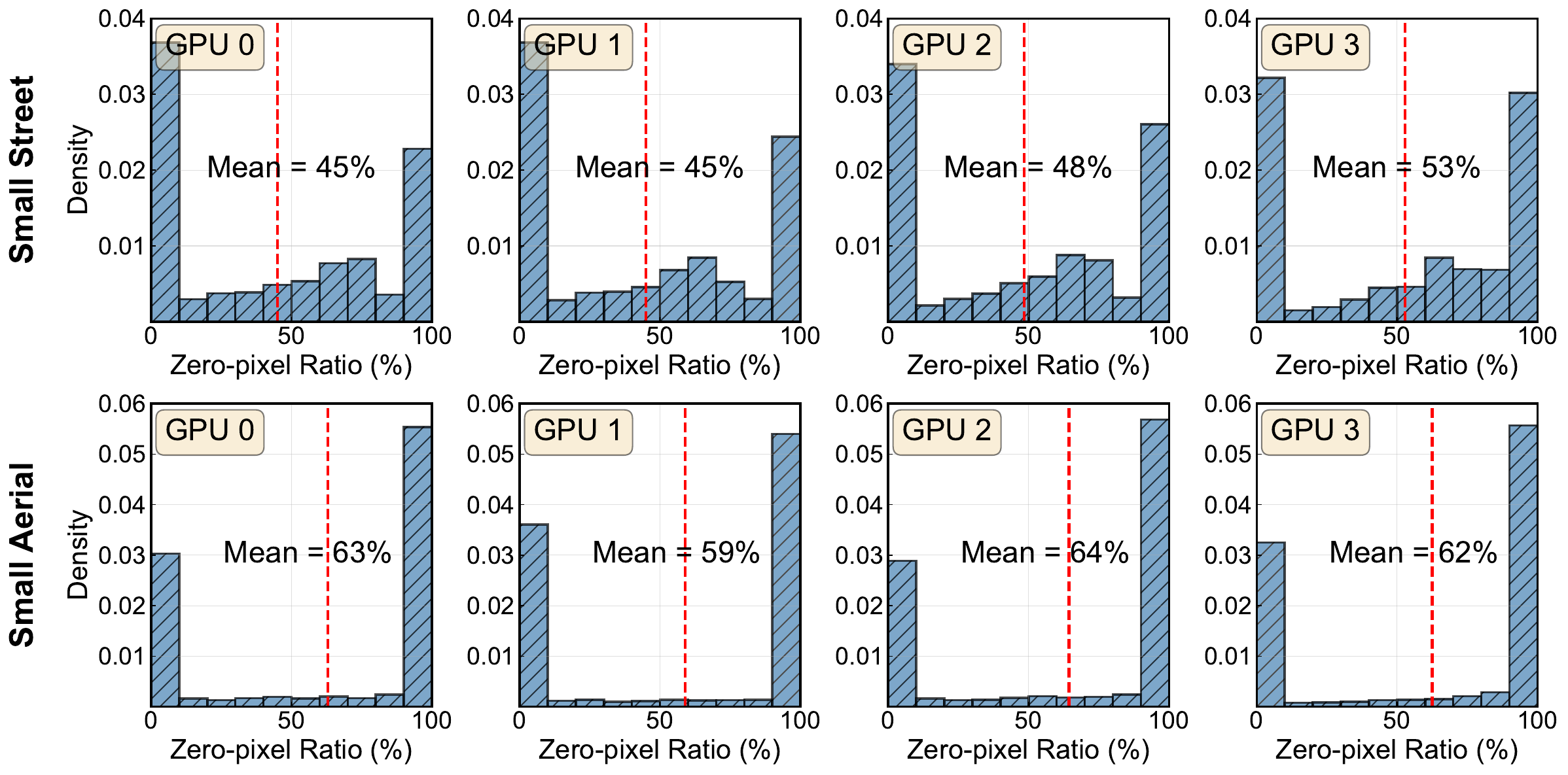}
    \vspace{-6.5mm}
    \caption{Zero-pixel ratio in naive all-gather communication.}
    \Description{Zero-pixel redundancy in naive all-gather communication.}
    \label{fig:design-zero-redundancy}
\end{subfigure}\hfill
\begin{subfigure}[t]{0.48\linewidth}
    \centering
    \includegraphics[width=\linewidth]{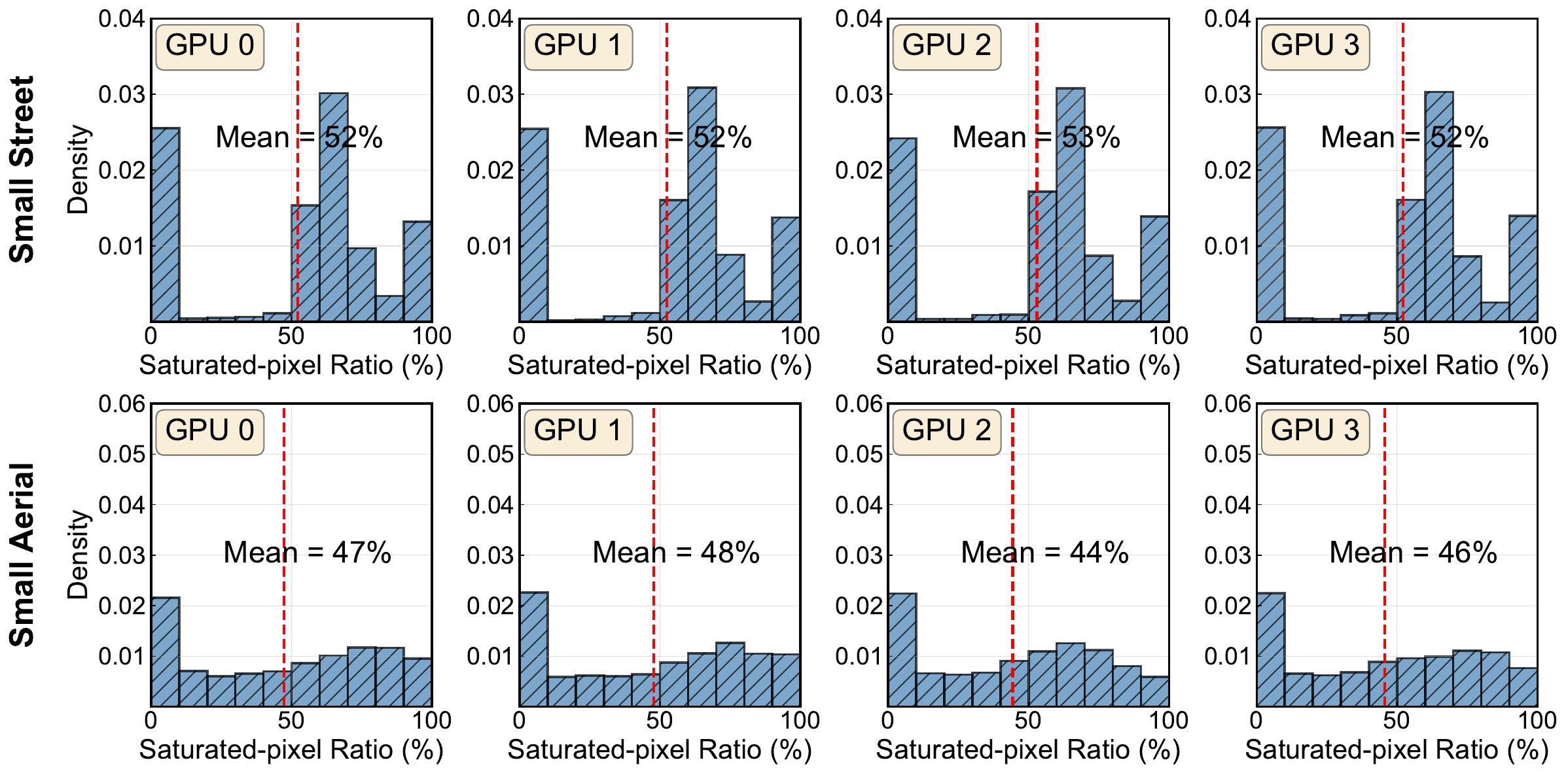}
    \vspace{-6.5mm}
    \caption{Saturated-pixel ratio with spatial redundancy removed.}
    \Description{Saturation redundancy after removing spatial redundancy.}
    \label{fig:design-trans-redundancy}
\end{subfigure}
\vspace{-3.5mm}
\caption{Redundancy analysis of pixel communication.  
(a) Over half of the transmitted pixels are zeros in the naive all-gather, showing dominant spatial redundancy. 
(b) With spatial redundancy removed, still many pixels remain saturated (transmittance $\approx 0$) with no contribution to final blending, indicating saturation redundancy. }
\label{fig:design-redundancy}
\end{figure*}

\begin{figure}[t]
\centering
\includegraphics[width=1.0\linewidth]{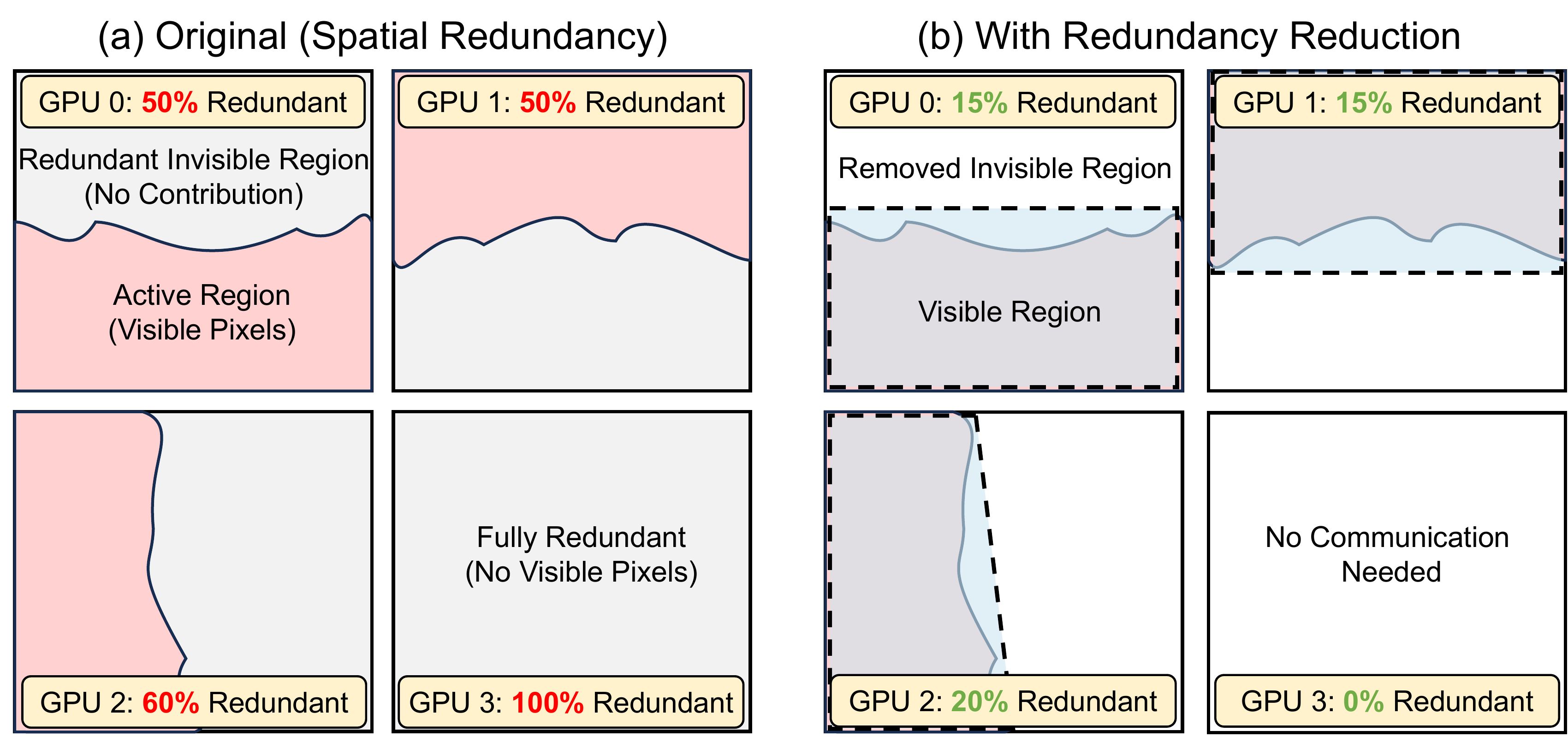}
\vspace{-7.5mm}
\caption{Spatial redundancy reduction.
(a) The original communication exhibits redundancy from invisible pixels.
(b) It eliminates invisible regions with redundant pixels.}
\Description{An illustration of reducing spatial redundancy by eliminating invisible pixels from communication.}
\label{fig:design-predict-pixel-range}
\vspace{-1mm}
\end{figure}

\subsection{Pixel-level Redundancy Reduction}
\label{sec:design-redundancy-reduction}

\noindent\ul{\textbf{Challenge: Pixel-level Redundancy from Geometric Invisibility and Transmittance Saturation.}}
Under the pixel-level communication scheme, we observe that a substantial portion of transmitted pixels never influence the final image, indicating significant bandwidth waste (Fig.~\ref{fig:design-redundancy}). These redundancies arise from two aspects.

\ul{\textit{i) Spatial redundancy from geometric invisibility.}} 
As illustrated in Fig.~\ref{fig:design-spatial-redundancy}a, each GPU covers only a portion of the scene 3D space, leaving large parts of the camera frustum outside its spatial geometry.
These invisible regions lead to substantial spatial redundancy (as demonstrated in Fig.~\ref{fig:design-spatial-redundancy}b and Fig.~\ref{fig:design-predict-pixel-range}a) during communication, where partial pixels from those unseen regions have no contributions to the final image rendering.
Consistently, Fig.~\ref{fig:design-redundancy}a shows that up to 64\% of transmitted pixels are zeros, further confirming the severity of this spatial redundancy.
This spatial redundancy arises from the geometric mismatch between the camera frustum and GPU partitions: if a region lies outside the frustum, it remains invisible and thus redundant to the training. 
That said, pixel-level communication needs to consider view-dependent visibility for partial pixel transmission.

\ul{\textit{ii) Saturation redundancy from transmittance invisibility.}}
Another form of redundancy emerges dynamically during training, as certain Gaussians become opaque and begin to behave like solid surfaces, such as walls and buildings.
As illustrated in Fig.~\ref{fig:design-spatial-redundancy}c, once earlier Gaussians along a camera ray have already reduced a pixel’s transmittance to nearly zero, subsequent Gaussians in the corresponding partitions on other GPUs no longer contribute to blending, yet still transmit their partial pixels that have already been saturated.
Fig.~\ref{fig:design-redundancy}b shows that even with spatial redundancy removed, up to 53\% of transmitted pixels exhibit transmittance close to zero. This observation highlights a significant challenge of redundant communication cost due to dynamic transmittance invisibility during the runtime training process.

\noindent\ul{\textbf{Opportunity: Predictable and Detectable Invisibility for Redundancy Reduction.}}
Since spatial and saturation redundancy arise from geometric and transmittance \textit{invisibility}, we identify the opportunity that invisibility can be \textit{predicted} or \textit{detected} to determine redundant pixel-level communication.

i) A convex region can be identified through its boundary. %
As illustrated in Fig.~\ref{fig:design-spatial-redundancy}a, the camera frustum is a convex polyhedron defined by its eight corner vertices, and each GPU partition is also convex, designed explicitly as an axis-aligned bounding box.
Since the intersection of two convex sets remains convex~\cite{boyd2004convex, bertsekas2003convex, hiriart2013convex}, the overlapping volume between a frustum and partition, e.g., ABCH in Fig.~\ref{fig:design-spatial-redundancy}a, is also a convex polyhedron representing part of the scene \textit{i.e., geometrically visible}.
Projecting this convex 3D space onto the image plane yields a convex visible region, e.g., ABCH in Fig.~\ref{fig:design-spatial-redundancy}b.
This convex region can be efficiently determined by projecting only the intersection vertices, as shown by ABCH, HCDE, GHEF in Fig.~\ref{fig:design-spatial-redundancy}b, without rendering.

ii) As the opacity of each Gaussian is dynamically updated during training, transmittance invisibility can be detected at runtime.
For a given camera view, the participating GPUs exchange partial pixel transmittance and depth information in a predefined order for global compositing, as described in Section~\ref{sec:design-pixel-level}.
During compositing, each GPU can locally evaluate the accumulated transmittance along each ray.
From the first GPU to the current GPU $m$, once the cumulative transmittance $\prod_{k \le m} T^k$ is less than a threshold $\varepsilon$, all subsequent partitions behind GPU $m$ have no meaningful contribution and thus become invisible on the camera ray.

\noindent\ul{\textbf{Design of Pixel-level Redundancy Reduction.}} To systematically remove the pixel-level communication redundancy, we propose a reduction algorithm that identifies the required GPUs and invisible pixels for a given camera view.

\begin{figure*}[t]
\centering
\includegraphics[width=1.0\linewidth]{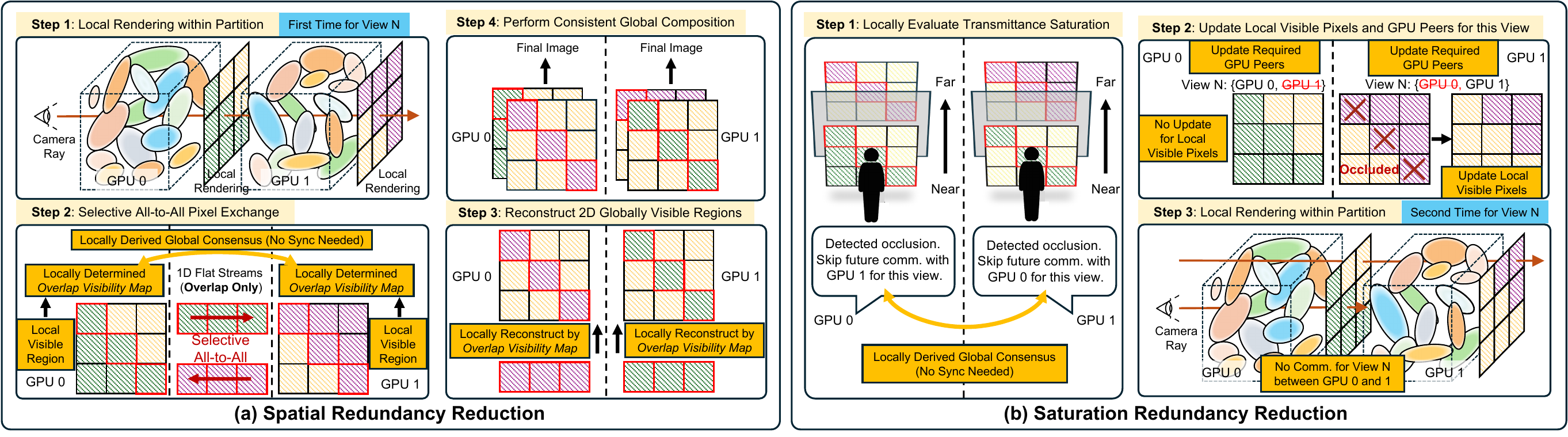}
\vspace{-8mm}
\caption{Distributed rendering with pixel-level redundancy reduction.
(a) After local rendering, \text{\abbr} identifies overlap visible pixel regions between GPUs based on geometric visibility, transmitting those pixels with all-to-all communication. 
(b) During training, \text{\abbr} dynamically detects transmittance saturation, and each GPU then renders only the remaining visible pixels according to its updated pixel set, effectively removing saturation redundancy caused by transmittance invisibility.}
\Description{Distributed rendering with pixel-level redundancy reduction.}
\label{fig:design-redundancy-reduction-workflow}
\vspace{-4mm}
\end{figure*}

\textit{\Circled{1} Spatial Redundancy Reduction.}  
\text{\abbr} leverages static geometric correlations between the camera frustum and GPU partition at initialization stage to identify pixel-level visibility.  
Each GPU is tied to a partition that consists of Gaussians belonging to the overlapping 3D geometric space within the frustum. The vertices of the 3D space are then projected onto the 2D image plane to obtain the visible pixel region for a given camera view, as shown in Fig.~\ref{fig:design-predict-pixel-range}b. Then, this visible region map can be shared across all GPUs to identify valid pixels without extra communication.
Thus, pixel transmission is restricted to the overlapping regions. The pixels are transmitted as compact streams via \textit{selective all-to-all} communication among GPUs that share visible regions.

Fig.~\ref{fig:design-redundancy-reduction-workflow}a illustrates the workflow of spatial redundancy reduction. At step 1, each GPU performs local rendering within its partition (convex partition according to the number of Gaussians).  
At step 2, the local visible regions are identified, i.e., $\Omega_v^m$ on GPU $m$ for the camera view $v$, which is used to determine the overlap visibility map between two GPUs. The local rendered pixels, whose corresponding locations on this map are valid, are then transmitted (selective all-to-all) by a flat 1D stream.  
At step 3, each GPU reconstructs the received 1D streams into its local 2D overlapping region map using identified pixel regions, ensuring spatial alignment.  
Finally, the global composition is performed to render the image, and the loss of visible regions is backpropagated to the corresponding GPUs to update the Gaussians.

\begin{figure}[t]
\centering
\includegraphics[width=1.0\linewidth]{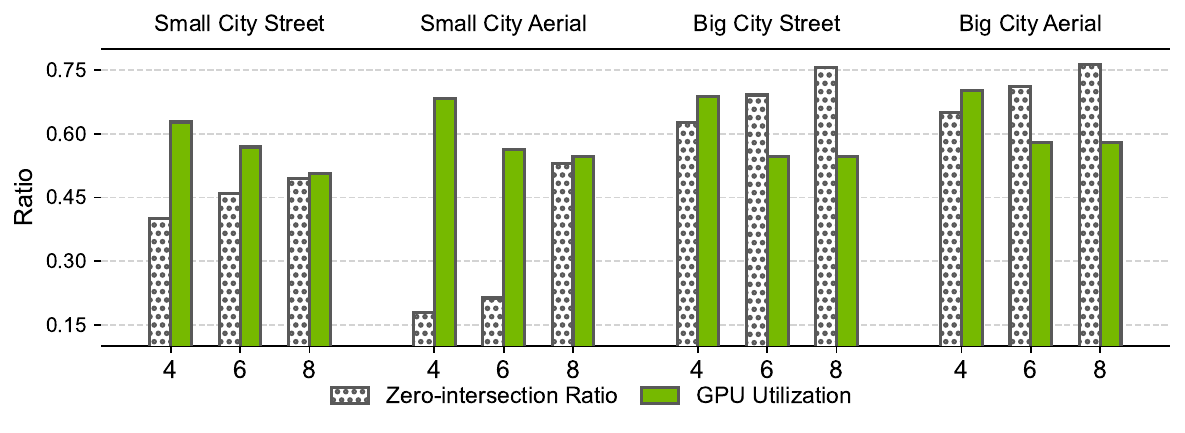}
\vspace{-9mm}
\caption{GPU utilization and zero-intersection ratio under the one-view-per-iteration strategy. As the number of GPUs increases, zero-intersection ratios of views rise while GPU utilization consistently drops, creating opportunities for efficient scheduling with conflict-free view consolidation.}
\Description{GPU utilization and zero-intersection ratio under the one-view-per-iteration strategy as the number of GPUs increases.}
\label{fig:scheduler-util-comp}
\vspace{-1mm}
\end{figure}

\textit{\Circled{2} Saturation Redundancy Reduction.}
\text{\abbr} leverages information from global composition to detect transmittance saturation during runtime training at no additional cost.  
Each GPU, while compositing colors from preceding partitions, determines which pixels are saturated by evaluating whether the cumulative transmittance is less than the threshold $\varepsilon$.
Once such saturation is detected, the affected pixels are removed from the local visible region~$\Omega_v^m$, 
and GPUs participating in the reconstruction skip further communication for these pixels in subsequent iterations of the same view.
As shown in Fig.~\ref{fig:design-redundancy-reduction-workflow}b,
GPU~1, after compositing with GPU~0, observes that three overlapping pixels have already saturated in pixels of GPU~0. GPU~1 locally identifies that GPU~0 is no longer a required peer for this view.
In the following iteration, each GPU renders only the remaining visible pixels according to its updated pixel set, while saturated regions are completely excluded from both rendering and supervision.
Over time, \text{\abbr} adaptively removes redundant communication and computation, effectively removing saturation redundancy by identifying transmittance invisibility.

\subsection{Conflict-free View Consolidation for Efficient Scheduling} 
\label{sec:design-scheduling}

\begin{figure}[t]
\centering
\includegraphics[width=1.0\linewidth]{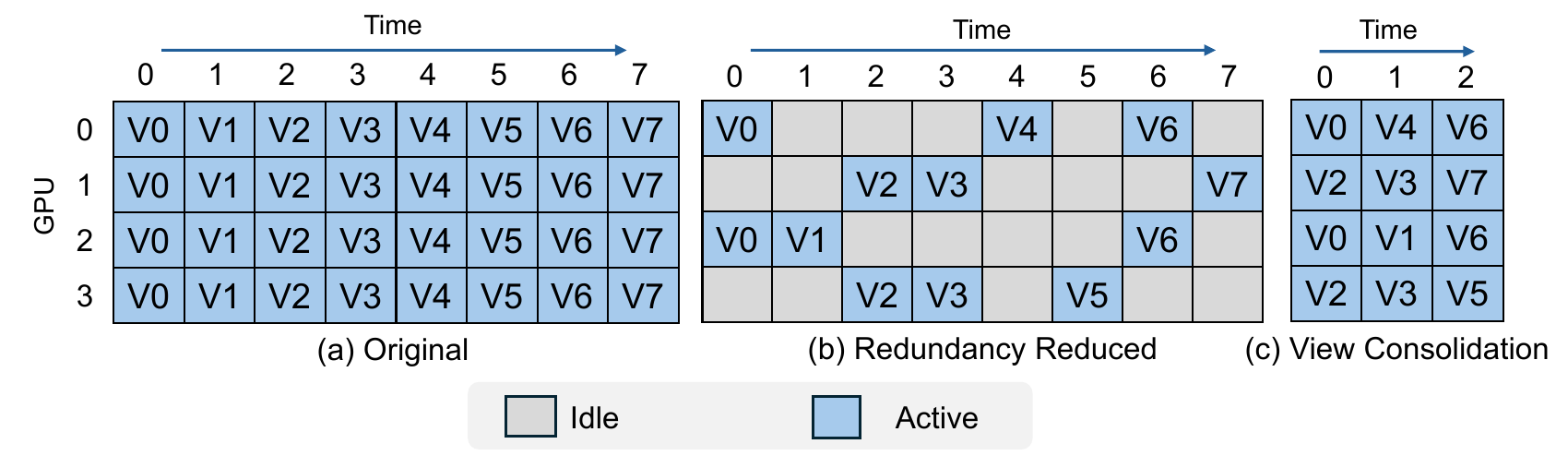}
\vspace{-8mm}
\caption{Illustration of view and GPU scheduling.}
\Description{Illustration of view and GPU scheduling.}
\label{fig:design-scheduling-bubble}
\vspace{-4mm}
\end{figure}

\noindent\ul{\textbf{Challenge: Insufficient GPU Resource Utilization.}}
The existing scheduling strategy takes one view for training each time on all GPUs due to the lack of view-GPUs mapping, as illustrated in Fig.~\ref{fig:design-scheduling-bubble}a. With our proposed redundancy reduction (Section~\ref{sec:design-redundancy-reduction}), even though redundant pixel-level communication can be removed, those GPUs may still remain idle and wait for active GPUs to complete the current view.
To analyze the GPU utilization~\cite{darabi2022morpheus, yang2023betty}, we define utilization ratio $U$ by measuring the average ratio of active GPUs for a training epoch, i.e., $U = |\mathcal{A}|/{ M}$, where $\mathcal{A}$ and $M$ denote the set of active GPUs and the total number of GPUs, respectively. Fig. \ref{fig:scheduler-util-comp} shows that $U$ is near 50\% in the Big City Street and Aerial datasets under the original one-view-per-iteration execution strategy, indicating significant insufficient GPU resource utilization. Moreover, the GPU utilization problem becomes more severe as the number of GPUs increases or the scene grows larger. 
This pervasive underutilization poses a key efficiency challenge as training scales.

\begin{figure}[t]
\centering
\includegraphics[width=1.0\linewidth]{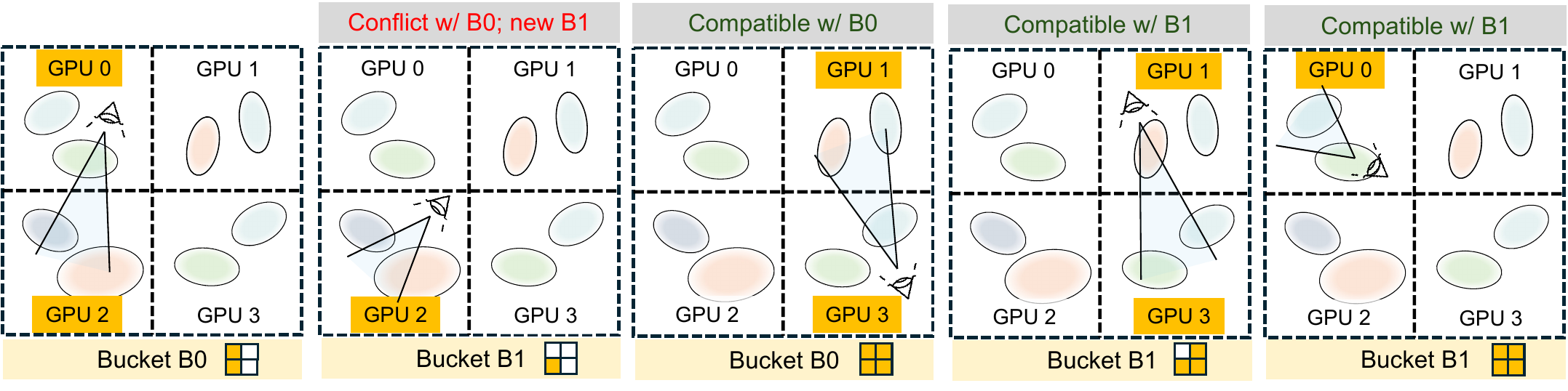}
\vspace{-7.5mm}
\caption{Demonstration of conflict-free view consolidation. Non-conflicting views are grouped to utilize GPUs fully, while a conflicting one will be assigned a new bucket.}
\Description{Demonstration of conflict-free view consolidation.}
\label{fig:design-view-bucketing-conflict}
\vspace{-2mm}
\end{figure}

\begin{table*}[t]
\caption{Overall training performance and quantitative quality with 8 GPUs on the large-scale datasets, in comparison with Grendel. Small City Street, Small City Aerial, Big City Street, and Big City Aerial are trained for 150k, 30k, 1.2M, and 300k iterations, respectively, for both methods.}
\vspace{-3.5mm}
\centering
\small
\resizebox{\textwidth}{!}{%
\begin{tabular}{l
    >{\centering\arraybackslash}p{1.6cm} >{\centering\arraybackslash}p{0.7cm} >{\centering\arraybackslash}p{0.7cm}
    >{\centering\arraybackslash}p{1.6cm} >{\centering\arraybackslash}p{0.7cm} >{\centering\arraybackslash}p{0.7cm}
    >{\centering\arraybackslash}p{1.6cm} >{\centering\arraybackslash}p{0.7cm} >{\centering\arraybackslash}p{0.7cm}
    >{\centering\arraybackslash}p{1.6cm} >{\centering\arraybackslash}p{0.7cm} >{\centering\arraybackslash}p{0.7cm}}
\toprule
\multirow{2}{*}{Method}
& \multicolumn{3}{c}{Small City Street}
& \multicolumn{3}{c}{Small City Aerial}
& \multicolumn{3}{c}{Big City Street}
& \multicolumn{3}{c}{Big City Aerial} \\
\cmidrule(lr){2-4} \cmidrule(lr){5-7} \cmidrule(lr){8-10} \cmidrule(lr){11-13}
& \# Gaussian & PSNR$\uparrow$ & Time$\downarrow$ 
& \# Gaussian & PSNR$\uparrow$ & Time$\downarrow$
& \# Gaussian & PSNR$\uparrow$ & Time$\downarrow$
& \# Gaussian & PSNR$\uparrow$ & Time$\downarrow$ \\
\midrule

Grendel     
& 55M   & 21.4 & 7.5h
& 37M   & 26.6 & 25.5m
& 120M  & \textbf{23.2} & 48.8h
& 120M  & 30.4 & 3.6h \\
\rowcolor{blue!12}
\textbf{\text{\abbr}}
& 55M   & \textbf{21.7} &  \textbf{1.3h}   
& 37M   & \textbf{27.3} &  \textbf{8.0m}            
& 120M  & \textbf{23.2}  & \textbf{6.4h}           
& 120M  & \textbf{31.2} & \textbf{1.2h} \\                               
\bottomrule
\end{tabular}
}
\label{tab:eval-train-perf}
\end{table*}

\begin{table}[t]
\centering
\vspace{-1mm}
\caption{Summary of the large-scale datasets.}
\vspace{-4mm}
\label{tab:eval-dataset-info}
\begin{tabular}{lccc}
\toprule
\textbf{Dataset} & \textbf{\# Train} & \textbf{\# Test} & \textbf{Area} \\
\midrule
Small City Aerial   & 5{,}621     & 741         & $\text{2.7 km}^2$    \\
Small City Street   & 27{,}505    & 3{,}256     & $\text{2.7 km}^2$    \\
Big City Aerial     & 51{,}632    & 9{,}066     & $\text{25.3 km}^2$   \\
Big City Street     & 206{,}412   & 22{,}744    & $\text{25.3 km}^2$   \\
\bottomrule
\end{tabular}
\vspace{-2mm}
\end{table}

\noindent\ul{\textbf{Opportunity: High Zero-intersection Ratio of Views as Scale Up.}} 
To reduce the idle bubble, we can merge two views into a single time slot if the GPUs they cross have no dependence. For example, in Fig.~\ref{fig:design-scheduling-bubble}, view V0 needs GPU 0 and 2 while view V2 needs GPU 1 and 3, which have no dependence. 
To quantitatively analyze the potential of removing the idle GPU bubble, we define the zero-intersection ratio of views as the proportion of views that have no intersection with other views among the entire dataset. 
As demonstrated in Fig.~\ref{fig:scheduler-util-comp}, the zero-intersection ratio is higher than 50\% when using 8 GPUs, indicating that a high proportion of camera views can be scheduled simultaneously. This observation presents a significant opportunity for conflict-free view consolidation, enabling efficient view and GPU scheduling.

\noindent\ul{\textbf{Design of Efficient Scheduling with Conflict-free View Consolidation.}} 
The key design of the scheduler is to iterate over all camera views to greedily and incrementally bucket more views that have no overlap with the GPUs in the existing buckets. For the camera view \(v\), the scheduler determines (using \texttt{GetParticipants(v)} by our redundancy reduction approach) the set of GPUs $\mathcal{P}_v$ required to process that view. Then, it searches the existing buckets to find one whose current GPU assignment has no conflict with $\mathcal{P}_v$. If a satisfied bucket exists, the view is inserted into it, and the bucket's GPU assignment is updated to include \(\mathcal{P}_v\). If no compatible bucket is found, a new bucket is created containing only view $v$. Finally, it returns the list of buckets that define the conflict-free execution scheduling for the current epoch. Fig. \ref{fig:design-view-bucketing-conflict} demonstrates our conflict-free view bucketing, with which the views can be efficiently scheduled by being processed concurrently within a bucket.  %
We also employ an asynchronous execution pipeline that overlaps communication and computation for view rendering and backpropagation, using a FIFO queue of in-flight views.

%% file: sec/5_evaluation.tex
\section{Performance and Scalability Evaluation}
\label{sec:eval}

\begin{figure}[t]
\centering
\includegraphics[width=1.0\linewidth]{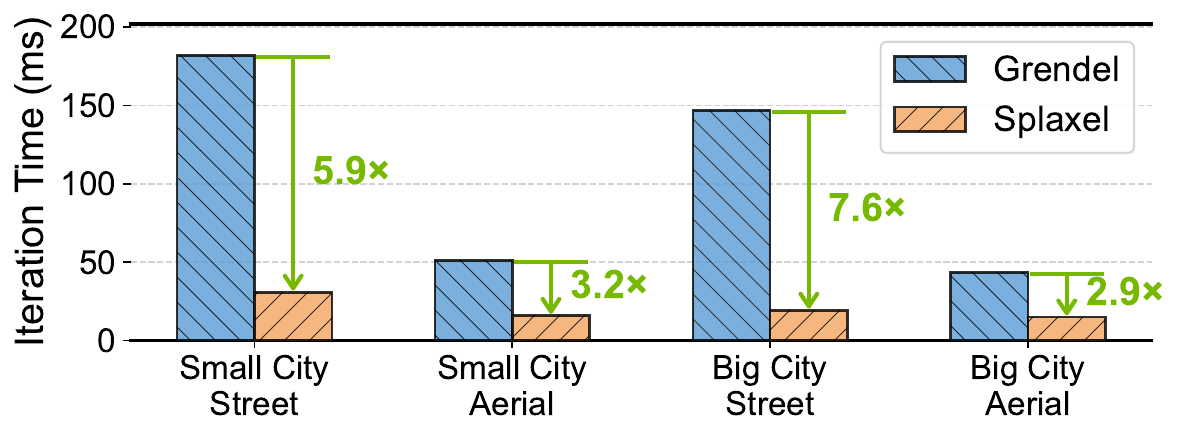}
\vspace{-9mm}
\caption{Per-iteration training (one view processed per iteration) time with 8 GPUs.}
\Description{Per-iteration training (one view processed per iteration) time with 8 GPUs.}
\label{fig:eval-dataset-times}
\vspace{-2mm}
\end{figure}

\begin{figure}[t]
\centering
\includegraphics[width=1.0\linewidth]{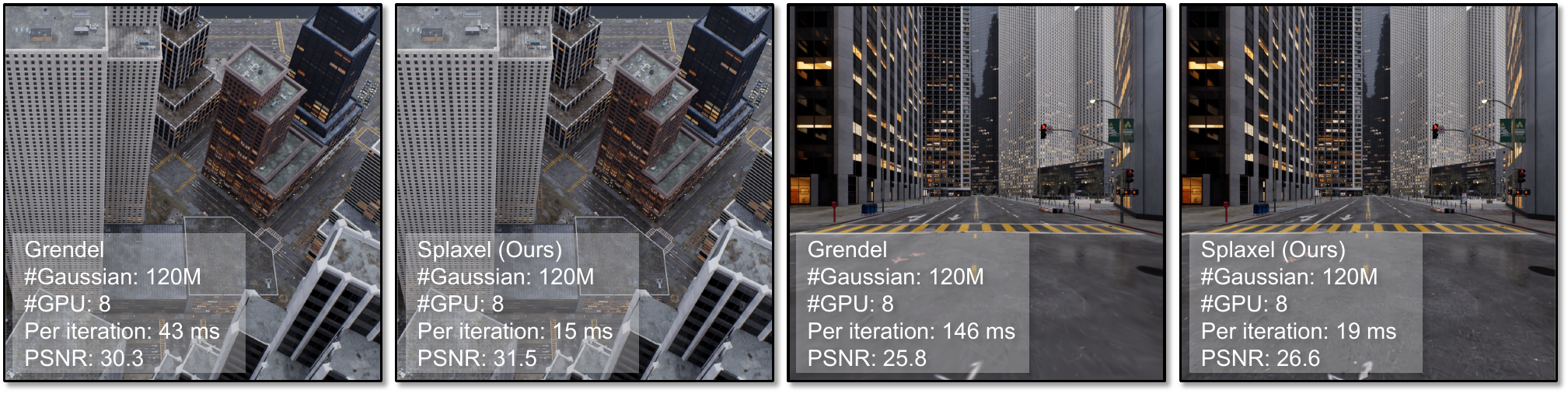}
\vspace{-8mm}
\caption{Example rendered images for visual reconstruction quality comparison. (Left) Big City Aerial; (Right) Big City Street. With the same 120M Gaussians, \text{\abbr} achieves higher rendering quality with much less training time. }
\Description{Example rendered images for visual reconstruction quality comparison.}
\label{fig:eval-img-comparison}
\vspace{-2mm}
\end{figure}

\noindent\ul{\textbf{Platform.}} We conduct evaluations on two platforms, each equipped with dual AMD EPYC 9254 24-core processors. Platform 1 is equipped with eight NVIDIA RTX 6000 Ada Generation GPUs (48 GB GDDR6), and Platform 2 has eight NVIDIA RTX PRO 6000 Blackwell GPUs (96 GB GDDR7).

\noindent\ul{\textbf{Datasets.}} We evaluate \text{\abbr} with four datasets from the \textit{MatrixCity}~\cite{li2023matrixcity} benchmark: \textit{Small City Street}, \textit{Small City Aerial}, \textit{Big City Street}, and \textit{Big City Aerial}, at 1080p resolution. The basic information of those datasets is summarized in Table~\ref{tab:eval-dataset-info}. Each dataset represents a distinct large-scale urban reconstruction scenario captured from either ground-level or aerial perspectives. Together, they cover a total area of approximately 28km$^{2}$, containing about 60K aerial images and 350K street-view images. We run the Small City experiments on Platform 1 and the Big City experiments on Platform 2. To the best of our knowledge, \textit{MatrixCity} remains the only publicly available benchmark of sufficient scale to support large-scale 3D scene reconstruction.

\begin{figure*}[t]
\centering
\includegraphics[width=\textwidth]{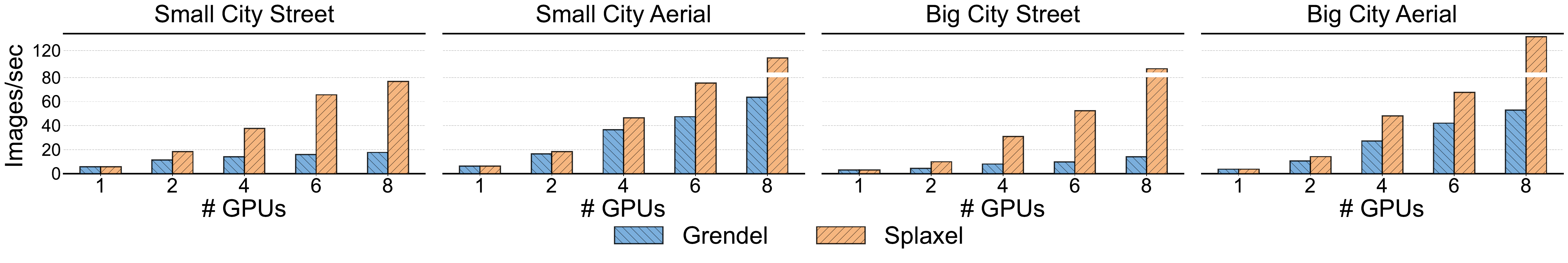}
\vspace{-9.5mm}
\caption{The training throughput of \text{\abbr} compared to the baseline, Grendel. As the number of GPUs increases from 1 to 8, the throughput of \text{\abbr} improves significantly, while Grendel scales with a marginal speedup. }
\Description{Training throughput of \text{\abbr} and Grendel as the number of GPUs increases from 1 to 8.}
\label{fig:eval-scalability}
\vspace{-4mm}
\end{figure*}

\noindent\ul{\textbf{Baseline \& Other Settings.}} We choose Grendel~\cite{zhao2025scaling3dgs} as our baseline. For a fair comparison, we follow Grendel to set the batch size equal to the number of GPUs.  We use the standard rendering quality metric, Peak Signal-to-Noise Ratio (PSNR), to evaluate the reconstruction quality. We use CUDA 12.8 on NVIDIA GPUs and PyTorch v2.9.0 for model training.

\subsection{Training Performance and Quality}

\begin{figure}[t]
\centering
\includegraphics[width=1.0\linewidth]{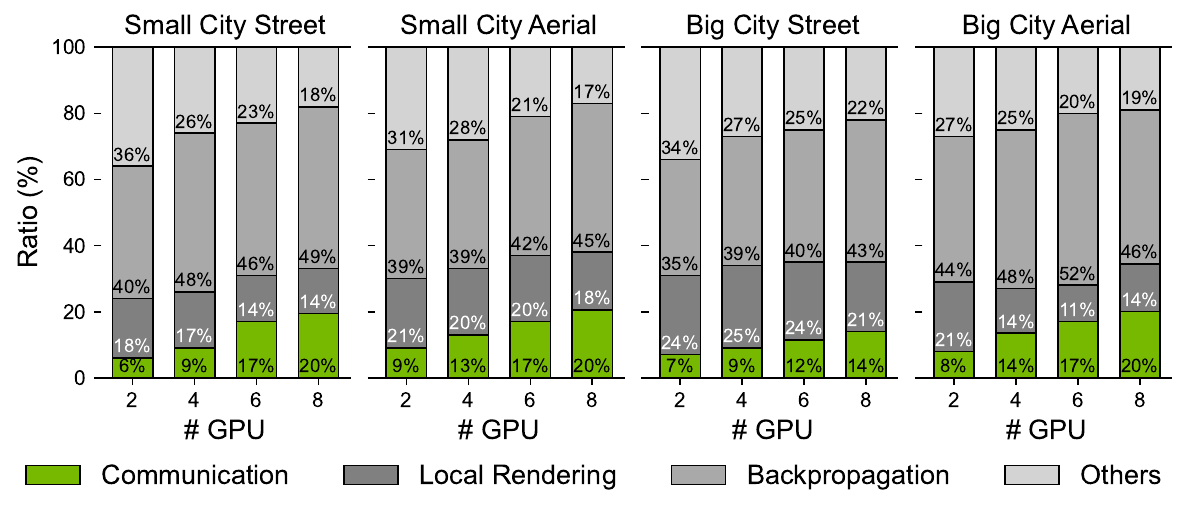}
\vspace{-9.5mm}
\caption{Per-iteration runtime breakdown over multiple GPUs. Communication is no longer a bottleneck.}
\Description{Per-iteration runtime breakdown over multiple GPU counts.}
\label{fig:eval-percentageBreakdown}
\vspace{-2mm}
\end{figure}

Table~\ref{tab:eval-train-perf} summarizes the overall training performance and reconstruction quality on the four datasets. Compared to Grendel, \text{\abbr} achieves significantly faster training with even better rendering quality. Fig.~\ref{fig:eval-dataset-times} shows the per-iteration time with 8 GPUs for each dataset and the corresponding speedups over Grendel. \text{\abbr} consistently reduces training time between 2.9$\times$ and 7.6$\times$ across all datasets. The most significant improvements occur on the street datasets, where iteration time decreases from 181.84ms to 30.82ms on Small City Street and from 146.77ms to 19.32ms on Big City Street. This is because \textit{street scenes} feature dense building facades, narrow corridors, and strong depth discontinuities. These structural characteristics introduce substantial redundancy, and many pixels have no geometric contribution for most partitions. \text{\abbr} can quickly identify these spatial and saturation redundancies. Fig.~\ref{fig:eval-img-comparison} shows representative scenes reconstructed from the Big City Aerial and Big City Street datasets trained on 8~GPUs, showing that \text{\abbr} achieves higher-quality reconstructions with significantly less training time, compared to Grendel~\cite{zhao2025scaling3dgs}.  %

\subsection{Scalability and Training Throughput}

Next, we evaluate the throughput of \text{\abbr} with different numbers of GPUs. To ensure \textit{a consistent problem setting} across different GPU counts, we choose 15M Gaussians for the Small City Street and Aerial scenes and 48M Gaussians for the Big City Street and Aerial scenes, which fit within single-GPU memory. Results with 120M Gaussians on 8 GPUs are provided in Table~\ref{tab:eval-train-perf}.
Fig.~\ref{fig:eval-scalability} reports the training throughput (views processed per second) across all four datasets. \text{\abbr} consistently achieves significantly higher throughput as the number of GPUs increases, reaching up to 90 images/s on Big City Street and 142 images/s on Big City Aerial with eight GPUs, compared to only 14 and 52 images/s for Grendel, respectively. In contrast, Grendel shows only modest gains beyond four GPUs on the Street datasets due to its communication-bound scaling. These results confirm that \text{\abbr} fully leverages multi-GPU parallelism, achieving consistent throughput gains as the GPU count scales.

Our results show that \text{\abbr} benefits more from larger scenes. As scene size increases, each GPU covers a larger region, while each camera view still only observes a subset of the scene. This reduces active GPUs per view and increases spatial redundancy, enabling \text{\abbr} to eliminate more invisible pixels and avoid redundant computation.

\begin{figure*}[t]
\centering
\begin{subfigure}[t]{0.48\linewidth}
    \centering
    \includegraphics[width=\linewidth]{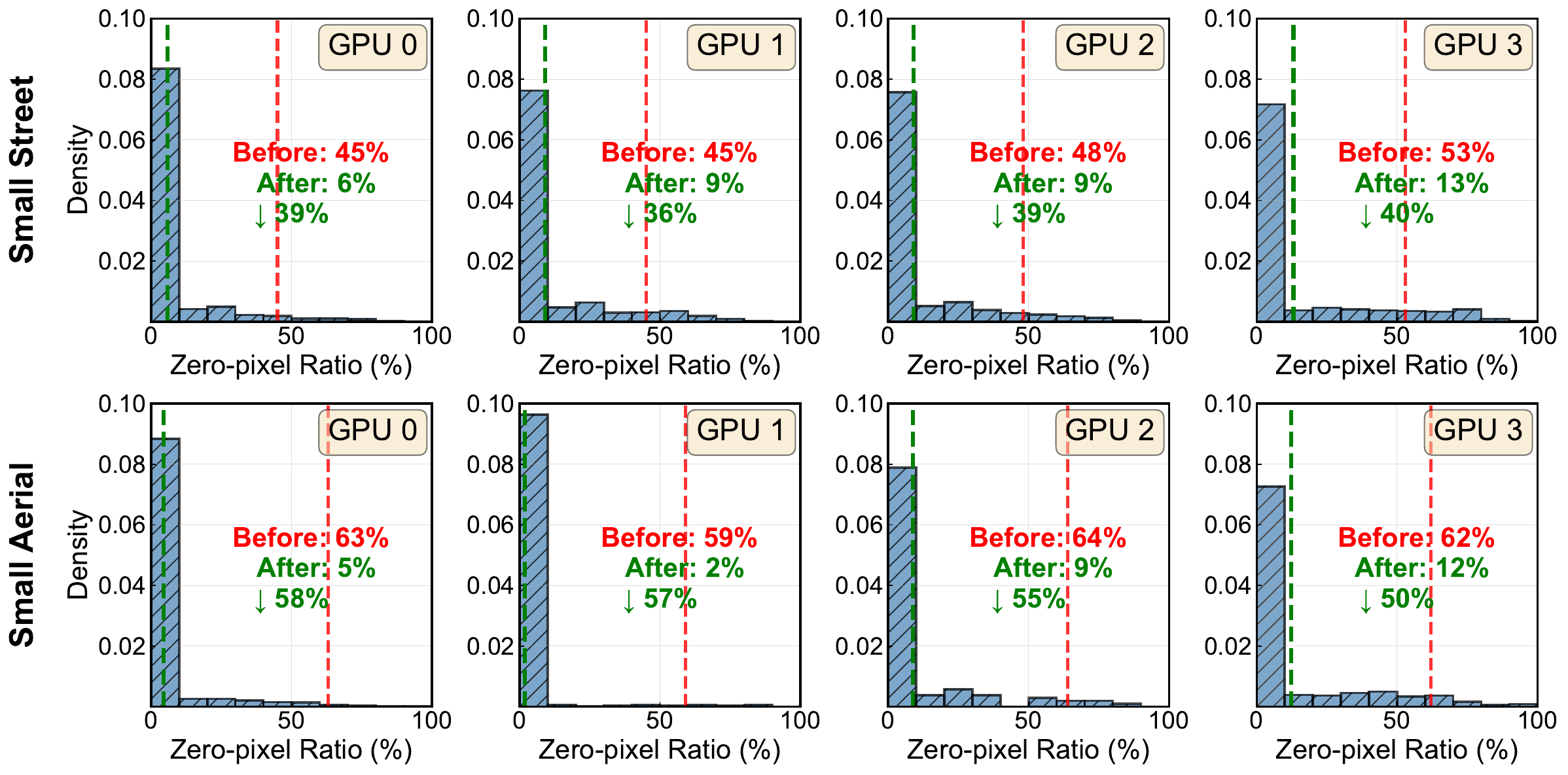}
    \vspace{-6.5mm}
    \caption{Zero-pixel ratio after removing spatial redundancy.}
    \label{fig:eval-zero-redundancy-remove}
\end{subfigure}\hfill
\begin{subfigure}[t]{0.48\linewidth}
    \centering
    \includegraphics[width=\linewidth]{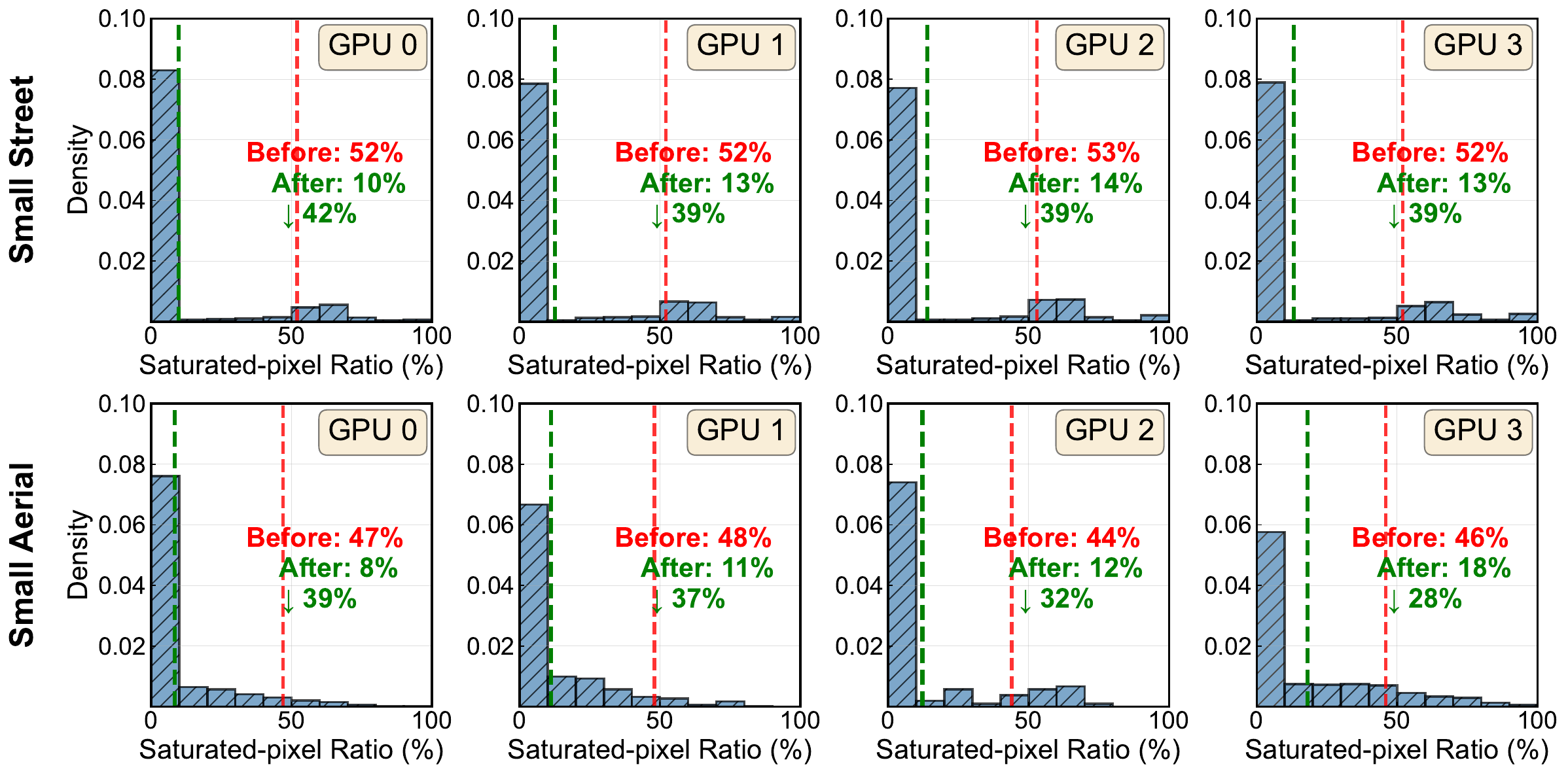}
    \vspace{-6.5mm}
    \caption{Saturated-pixel ratio after removing saturation redundancy.}
    \label{fig:eval-trans-redundancy-remove}
\end{subfigure}
\vspace{-4mm}
\caption{Quantitative illustration of reduced redundancy.}
\Description{Quantitative illustration of reduced redundancy.}
\label{fig:eval-redundancy-remove}
\vspace{-1mm}
\end{figure*}

\begin{figure}[t]
\centering
\includegraphics[width=\linewidth]{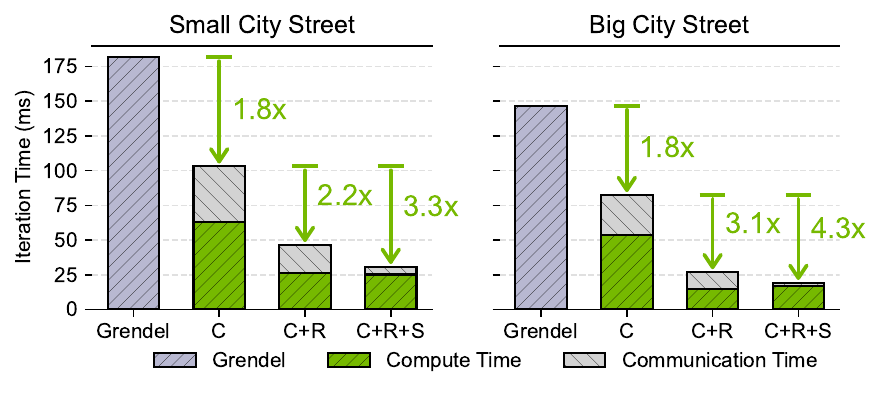}
\vspace{-9mm}
\caption{
Impact analysis of \text{\abbr} key design components.
C: pixel-level communication; R: redundancy reduction; S: view consolidation.}
\Description{Ablation study.}
\label{fig:eval-ablation}
\vspace{-1mm}
\end{figure}

\subsection{Per-iteration Runtime Breakdown}
\label{sec:eval-breakdown}

We provide breakdown results of the latency of communication, backpropagation, local rendering, and other components on the four large-scale datasets as the GPU count scales from 2 to 8, to systematically analyze the training time with \text{\abbr}.
From Fig.~\ref{fig:eval-percentageBreakdown}, across all four scenes, the breakdown shows that communication remains a small and stable fraction of the iteration time (roughly 10--20\%) even as the number of GPUs increases, confirming that communication is no longer the bottleneck under our pixel-level communication scheme. Local rendering consistently stays below 20\% and generally decreases with more GPUs, indicating that our method effectively eliminates redundant rendering work. As expected, backpropagation dominates the runtime across all configurations, reflecting its inherently heavy computation and demonstrating that \text{\abbr} successfully shifts the critical path back to the core learning step rather than communication or rendering overhead. Overall, the results validate the efficiency and scalability.

\subsection{Impact of Key Design Components}
\label{sec:eval-ablation}

To understand the contribution of each proposed component in \text{\abbr} to the final training acceleration, we evaluate three configurations:
i) pixel-level communication, where all GPUs participate in every view;
ii) pixel-level communication with redundancy reduction;
iii) pixel-level communication, redundancy reduction, and view consolidation. As shown in Fig.~\ref{fig:eval-ablation}, on Small City Street, pixel-level communication takes about 100ms per iteration.
With the redundancy reduction, it yields a 2.2$\times$ speedup.
Both communication and computation time decrease because far fewer GPUs participate in each view.
With the view consolidation, the speedup reaches 3.3 times over i). Compared to ii), computation time remains the same, while communication time drops significantly.
On Big City Street, pixel-level communication takes about 80ms per iteration. With further redundancy reduction, it achieves a 3.1$\times$ speedup over i). With view consolidation, the speedup increases to 4.3$\times$ over i). This ablation study demonstrates that each proposed component is efficient in improving the distributed training efficiency for large scenes.

\begin{figure}[t]
\centering
\includegraphics[width=0.95\linewidth]{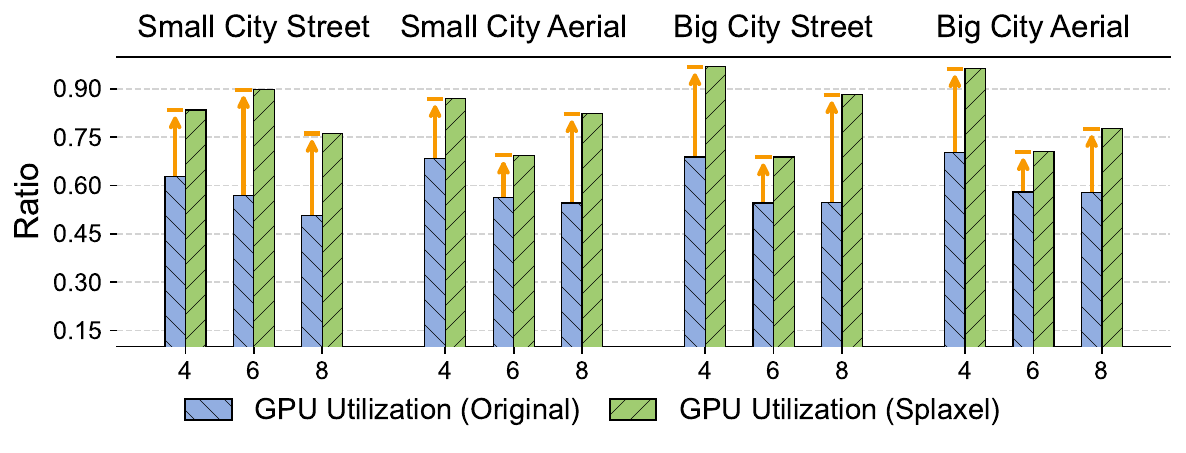}
\vspace{-5.5mm}
\caption{GPU utilization improvement.}
\Description{GPU utilization improvement.}
\label{fig:eval-util-improve}
\vspace{-1.5mm}
\end{figure}

\subsection{Technical Effectiveness and Analysis}
\label{sec:eval-discussion}

\ul{\textbf{Redundancy Reduction.}} Fig.~\ref{fig:eval-redundancy-remove} quantifies the reduced pixel-level redundancy across four distributed GPUs.
Fig.~\ref{fig:eval-redundancy-remove}a shows that by predicting geometric invisibility, frustum-invisible pixels are eliminated, decreasing the zero-pixel ratio by approximately 39\% on Small Street and 55\% on Small Aerial.
Fig.~\ref{fig:eval-redundancy-remove}b shows that during training, transmittance-invisible pixels are detected and then removed, lowering the saturated-pixel ratio by 40\% and 34\%, respectively.

\noindent\ul{\textbf{GPU Utilization Improvement.}} Fig.~\ref{fig:eval-util-improve} reports the GPU utilization across all four scenes as the number of GPUs increases. Our scheduler improves the GPU utilization by 20\%--35\% depending on the scene. Although utilization declines with increasing GPU count, our method yields significantly higher ratios of throughput. 

\begin{figure}[t]
\centering
\includegraphics[width=\linewidth]{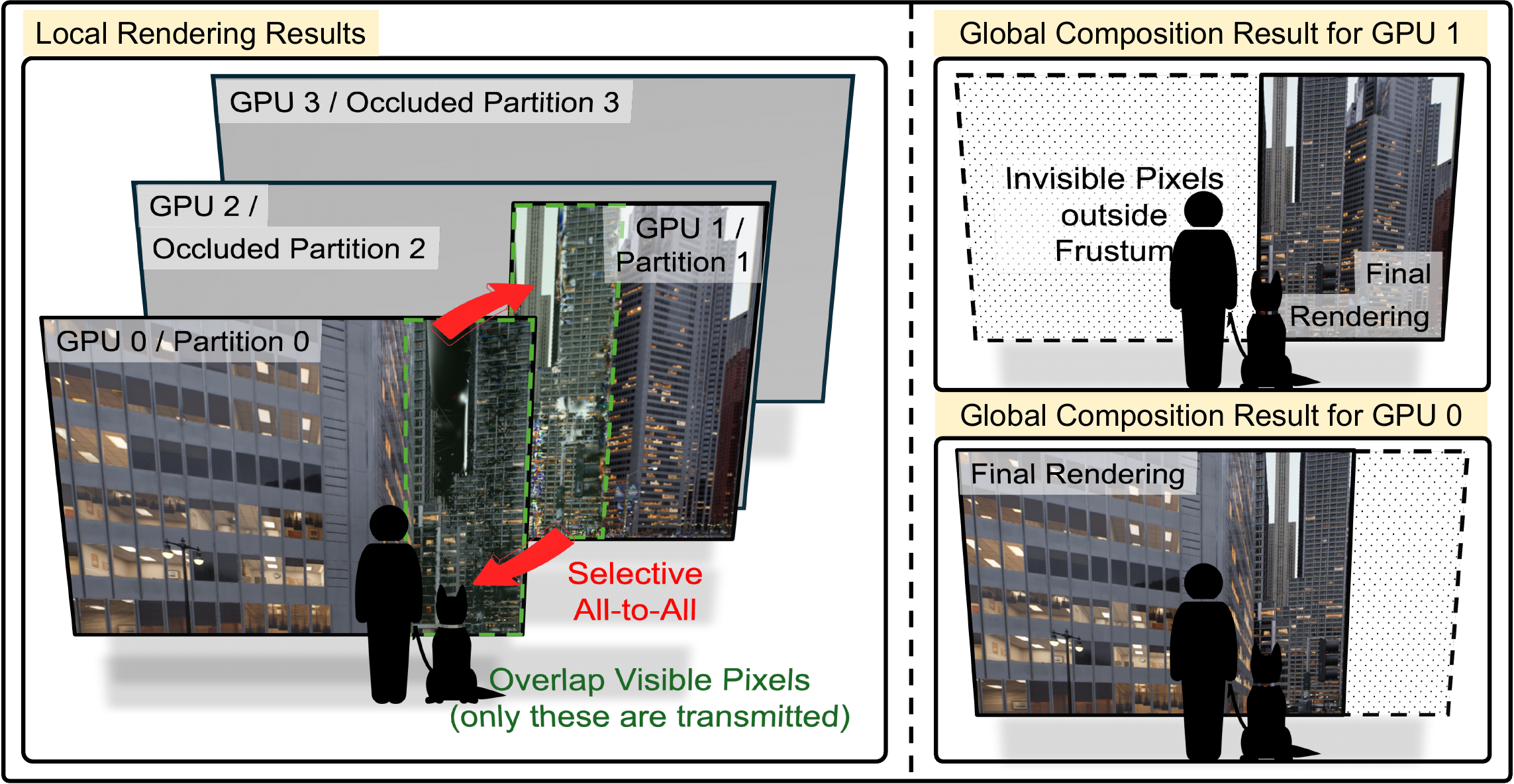}
\vspace{-7mm}
\caption{Real 3DGS rendering produced by \abbr. GPU 2 and GPU 3 are fully occluded and skipped, while GPU 0 and GPU 1 exchange overlap pixels (green). Artifacts appearing in the local overlap regions vanish after global composition.}
\Description{Real 3DGS rendering produced by \abbr.}
\label{fig:discuss-proxy}
\end{figure}

\ul{\textbf{Real-scene Illustration.}}
Fig.~\ref{fig:discuss-proxy} shows a real 3DGS rendering from \text{\abbr} during distributed training.
\text{\abbr} predicts per-GPU visible regions: GPU 0 covers most of the view, GPU 1 about half, while GPU 2 and GPU 3 nominally cover the full frame.
At runtime, however, the \text{\abbr} detects that all pixels become saturated after blending GPU 0 and GPU 1, so the contributions from GPU 2 and GPU 3 are skipped (grey panels).
Because each GPU locally derives its partition-to-pixel map, GPU 0 and GPU 1 precisely identify their overlap regions.
Only these overlap pixels (green) are exchanged via selective all-to-all, and artifacts appearing in the local overlap regions disappear after global composition, showing how \text{\abbr} restricts communication to pixels that truly matter.

\noindent\ul{\textbf{Batch Size.}}\label{reb-B5}
We evaluate different batch sizes for both \text{\abbr} and Grendel in Table~\ref{tab:reb-batch-size}. Larger batch sizes improve throughput for both due to reduced per-iteration overhead and improved GPU utilization. The speedup of \text{\abbr} over Grendel is up to $5.8$--$5.9\times$ and remains consistent across all settings.

\begin{table}[t]
\centering
\caption{Analysis of batch size. Larger batch sizes improve throughput for both methods by reducing per-iteration overhead. The speedup of \text{\abbr} over Grendel remains consistent across batch sizes.}
\vspace{-3mm}
\label{tab:reb-batch-size}
\footnotesize\footnotesize
\begin{tabular}{lccccc}
\toprule
\multirow{2}{*}[-2pt]{Batch Size}
& \multicolumn{2}{c}{Grendel}
& \multicolumn{2}{c}{\text{\abbr}}
& \multirow{2}{*}[-2pt]{Speedup} \\
\cmidrule(lr){2-3} \cmidrule(lr){4-5}
& PSNR$\uparrow$ & Time$\downarrow$
& PSNR$\uparrow$ & Time$\downarrow$ & \\
\midrule
4  & 21.45 & 8.2h & 21.74 & 1.41h & 5.8$\times$ \\
8  & 21.40 & 7.5h & 21.72 & 1.27h & 5.9$\times$ \\
16 & 21.16 & 7.0h & 21.19 & 1.18h & 5.9$\times$ \\
\bottomrule
\end{tabular}
\end{table}

\noindent\ul{\textbf{Sensitivity of Transmittance Threshold.}}
\label{reb-D1}
Our distributed rendering pipeline skips subsequent partitions once a pixel is considered saturated. Here, transmittance ranges from 1 for complete transparency to 0 for full saturation. The threshold therefore governs the trade-off between rendering efficiency and reconstruction quality. To study its effect, we evaluate different threshold values in Table~\ref{tab:reb-threshold-sensitivity}. Since PSNR is largely insensitive to the threshold, we use a fixed threshold of $1\times10^{-4}$ across all datasets without per-scene tuning.

\begin{table}[t]
\caption{Sensitivity to transmittance threshold. A fixed threshold of $1\times10^{-4}$ is used across all datasets. PSNR remains stable for thresholds $\leq 1\times10^{-2}$, confirming low sensitivity and no need for dataset-specific tuning.}
\vspace{-3mm}
\label{tab:reb-threshold-sensitivity}
\centering
\footnotesize
\begin{tabular}{lcccc}
\toprule
\multirow{2}{*}[0.85pt]{Threshold}
& \shortstack{Small City\\Street} & \shortstack{Small City\\Aerial} & \shortstack{Big City\\Street} & \shortstack{Big City\\Aerial} \\
\cmidrule(lr){2-2} \cmidrule(lr){3-3} \cmidrule(lr){4-4} \cmidrule(lr){5-5}
& PSNR$\uparrow$ & PSNR$\uparrow$ & PSNR$\uparrow$ & PSNR$\uparrow$ \\
\midrule
$1\times10^{-1}$ & 21.48 & 27.05 & 22.94 & 30.91 \\
$1\times10^{-2}$ & 21.65 & 27.22 & 23.14 & 31.16 \\
$1\times10^{-3}$ & 21.71 & 27.29 & 23.19 & 31.19 \\
\rowcolor{blue!12}
$\mathbf{1\times10^{-4}}$ & \textbf{21.72} & \textbf{27.30} & \textbf{23.20} & \textbf{31.23} \\
$1\times10^{-5}$ & 21.71 & 27.28 & 23.22 & 31.23 \\
\bottomrule
\end{tabular}
\end{table}

%% file: sec/6_related_work.tex
\section{Related Works}
\noindent\ul{\textbf{Acceleration on 3DGS Rendering. }}
Recent efforts focus on improving 3DGS rendering efficiency~\cite{hanson2025pup, hou2024sort, girish2024eagles, guedon2024sugar, wei2025no, lin2025metasapiens, li2025uni, yu2024mip}. LightGaussian~\cite{fan2024lightgaussian} reduces redundancy via pruning and feature compression, improving throughput with minimal quality loss. Compression-based methods~\cite{xie2024sizegs, yang2024spectrally, liu2025maskgaussian, niedermayr2024compressed, fang2024mini} further optimize storage and computation. 
On the system side, FlashGS~\cite{feng2025flashgs} improves software rasterization through efficient workload partitioning and pipelined tile-based rendering. Hardware accelerator designs~\cite{ye2025gaussian, streaminggs25dac, neo2026asplos, ags2026asplos} further boost on-device rendering by exploiting data reuse, memory-centric streaming, and inter-frame redundancy. Nebula~\cite{nebula2026asplos} scales rendering to large scenes via collaborative cloud-client processing with accelerated stereo rasterization.

\noindent\ul{\textbf{Acceleration on 3DGS Training. }}
3DGS training suffers from a large volume of atomic operations during gradient accumulation, and several works~\cite{durvasula2023distwar, mallick2024taming, liteGS, he2025gsarch, sun20243dgstream} have been proposed to alleviate this bottleneck. 
ARC~\cite{durvasula2025arc} accelerates atomic-intensive workloads by exploiting intra-warp locality for register-level reductions and by balancing atomic execution across hardware units, reducing global atomic traffic and improving throughput.
GSArch~\cite{he2025gsarch} introduces a customized hardware that speeds up 3DGS training by pruning gradients and reorganizing memory requests into disjoint patterns to mitigate contention. 

\noindent\ul{\textbf{Large-scale 3DGS Training. }}
Scaling 3DGS training has been explored through both algorithm- and system-level approaches~\cite{wu2024gauspu, hanson2025speedy, lee2024gscore, yu2024gaussian, zhang2025quadratic, xie2025generative}. Algorithm-level methods~\cite{su2025hug, liu2024citygaussianv2} typically adopt a divide-and-conquer strategy, partitioning scenes into smaller chunks for independent training and later merging. While effective for avoiding memory limits, these methods deviate from the original 3DGS pipeline and may degrade reconstruction quality. On the system side, GS-Scale~\cite{gsscale2026asplos} and CLM~\cite{clm2026asplos} overcome single-GPU memory limitations by offloading Gaussian data to host memory, enabling training of large-scale scenes on consumer-grade GPUs. However, these approaches scale memory capacity only and remain confined to a single GPU. Grendel~\cite{zhao2025scaling3dgs} is the only distributed multi-GPU framework that preserves global consistency by enabling training of tens of millions of Gaussians. However, its reliance on Gaussian-level communication leads to rapidly increasing data transmission overhead as scene size and GPU count grow, significantly limiting scalability.

%% file: sec/7_conclusion.tex
\section{Conclusion}
In this paper, we present \text{\abbr}, an efficient and principled distributed 3DGS training framework for large-scale scene reconstruction via pixel-level communication, achieving up to 7.6$\times$ speedup over the baseline.

%% file: sec/8_appendix.tex
\section{Appendix}
\subsection{Implementation Details}

\label{sec:implement-detail}

\begin{figure}[t]
\centering
\includegraphics[width=\linewidth]{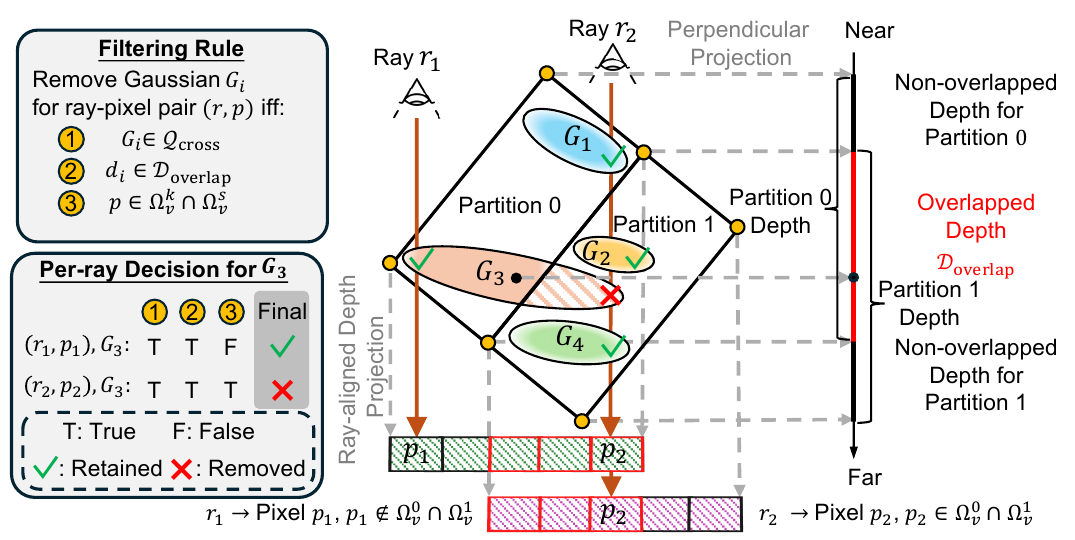}
\vspace{-8mm}
\caption{Demonstration of cross-boundary Gaussians handling in \text{\abbr}.}
\label{fig:reb-cross-boundary-handling}
\end{figure}

\ul{\textbf{Cross-boundary Gaussians Handling.}}\label{sec:reb-B1}
During training, Gaussian scales may expand across partition boundaries, which can break the global visibility order in distributed rendering and therefore require special handling to avoid incorrect alpha-composition and resulting PSNR degradation.
Since Gaussians are assigned to partitions by their mean positions, a Gaussian such as $G_3$ in Fig.~\ref{fig:reb-cross-boundary-handling} is assigned to partition~0 although its spatial support extends into partition~1.
During rendering, each partition independently composites its local Gaussians along each ray and communicates the resulting partial pixels for global composition.
This can introduce \emph{interleaving} in the global alpha-composition when multiple partitions contribute: for ray $r_2$, partition~0 composites $G_1\!-\!G_3$ while partition~1 composites $G_2\!-\!G_4$, yielding an interleaved global sequence $G_1\!-\!G_3\!-\!G_2\!-\!G_4$ that breaks correct occlusion and degrades PSNR, since alpha-composition is visibility-order dependent.

To avoid such interleaving, we apply \emph{per-ray filtering} to cross-boundary Gaussians.
For two adjacent partitions $k$ and $s$, let $\mathcal{Q}_\text{cross}$ denote the set of Gaussians whose spatial support spans both partitions.
We remove a Gaussian $G_i \in \mathcal{Q}_\text{cross}$ from the ray $r$ of pixel $p$ if and only if its depth $d_i$ falls within the overlapped depth interval $\mathcal{D}_{\text{overlap}}$ and $p$ lies in the overlapped pixel region $\Omega_v^k \cap \Omega_v^s$ between the visibility maps of partitions $k$ and $s$.
In Fig.~\ref{fig:reb-cross-boundary-handling}, $G_3 \in \mathcal{Q}_{\text{cross}}$ with $d_3 \in \mathcal{D}_{\text{overlap}}$ is therefore removed for ray $r_2$ (since $p_2 \in \Omega_v^0 \cap \Omega_v^1$) but retained for ray $r_1$ (since $p_1 \notin \Omega_v^0 \cap \Omega_v^1$), preserving consistent depth ordering.
The loss at $p_2$ is then minimized by the remaining Gaussians $G_1\!-\!G_2\!-\!G_4$: partition~0 composites $G_1$ and partition~1 composites $G_2\!-\!G_4$, yielding the consistent global ordering $G_1\!-\!G_2\!-\!G_4$ without interleaving, which provides geometry-aware regularization.

As shown in Table~\ref{tab:reb-cross-boundary-ablation}, per-ray filtering strategy improves PSNR by 0.2--0.4 dB across all scenes by avoiding inconsistent depth ordering from cross-boundary Gaussians. Moreover, the introduced training time overhead is negligible.

\noindent\ul{\textbf{Partition Strategy.}}\label{reb-A1-B2-B4}
Gaussians are assigned to partitions based on their mean positions, where the partition boundaries are determined via recursive binary splitting akin to a KD-tree~\cite{bentley1975multidimensional}, selecting the axis with the largest extent and splitting at the median Gaussian position at each level so that the Gaussian counts across partitions remain balanced.
The resulting axis-aligned boxes serve as convex partitions assigned to GPUs.
To evaluate sensitivity to partition imbalance, we deliberately create different levels of load imbalance, quantified by the imbalance ratio defined as $N_{\text{max}}/\bar{N} - 1$ where $N_{\text{max}}$ is the Gaussian count of the most loaded partition and $\bar{N}$ is the average across all partitions.
We disable densification and report average per-iteration time over sampled iterations.
As shown in Table~\ref{tab:reb-partition-sensitivity}, even at 20\% imbalance, \text{\abbr}'s per-iteration time increases by only 15--19\%, while still maintaining a $2.8$--$5.6\times$ speedup over Grendel, demonstrating that our KD-tree partitioning is robust to moderate load imbalance.

\begin{table}[t]
\centering
\caption{Partition sensitivity analysis. Per-iteration time under varying initial partition imbalance levels.}
\vspace{-4mm}
\label{tab:reb-partition-sensitivity}
\resizebox{\columnwidth}{!}{%
\begin{tabular}{lccccc}
\toprule
\multirow{2}{*}{Dataset} & Initial & \multicolumn{2}{c}{Avg.\ Iter.\ Time} & \multirow{2}{*}{Speedup} & Time Incr. \\
\cmidrule(lr){3-4}
 & Imbalance & Grendel & \text{\abbr} & & vs.\ 0\% \\
\midrule
\multirow{3}{*}{Small City Street}
 & 0\%  & 181.7ms & 30.1ms & 6.0$\times$ & ---   \\
 & 10\% & 188.6ms & 32.3ms & 5.8$\times$ & 7.3\%  \\
 & 20\% & 199.3ms & 35.3ms & 5.6$\times$ & 17.3\% \\
\midrule
\multirow{3}{*}{Small City Aerial}
 & 0\%  & 54.0ms  & 15.7ms & 3.4$\times$ & ---   \\
 & 10\% & 55.3ms  & 17.2ms & 3.2$\times$ & 9.6\%  \\
 & 20\% & 56.8ms  & 18.6ms & 3.1$\times$ & 18.5\% \\
\midrule
\multirow{3}{*}{Big City Street}
 & 0\%  & 150.1ms & 19.0ms & 7.9$\times$ & ---   \\
 & 10\% & 155.5ms & 20.1ms & 7.7$\times$ & 5.8\%  \\
 & 20\% & 158.4ms & 21.9ms & 7.2$\times$ & 15.3\% \\
\midrule
\multirow{3}{*}{Big City Aerial}
 & 0\%  & 46.8ms  & 15.4ms & 3.0$\times$ & ---   \\
 & 10\% & 48.7ms  & 16.3ms & 3.0$\times$ & 5.8\%  \\
 & 20\% & 51.3ms  & 18.4ms & 2.8$\times$ & 19.5\% \\
\bottomrule
\end{tabular}%
}
\end{table}

\begin{table*}[t]
\centering
\caption{Ablation on cross-boundary filtering. Enabling per-ray filtering improves PSNR by 0.2--0.4 dB across all scenes at negligible training time overhead.}
\vspace{-3mm}
\label{tab:reb-cross-boundary-ablation}
\footnotesize
\setlength{\tabcolsep}{10pt}
\begin{tabular}{lcccccccc}
\toprule
\multirow{2}{*}[-2pt]{Method}
& \multicolumn{2}{c}{Small City Street}
& \multicolumn{2}{c}{Small City Aerial}
& \multicolumn{2}{c}{Big City Street}
& \multicolumn{2}{c}{Big City Aerial} \\
\cmidrule(lr){2-3} \cmidrule(lr){4-5} \cmidrule(lr){6-7} \cmidrule(lr){8-9}
& PSNR$\uparrow$ & Time$\downarrow$
& PSNR$\uparrow$ & Time$\downarrow$
& PSNR$\uparrow$ & Time$\downarrow$
& PSNR$\uparrow$ & Time$\downarrow$ \\
\midrule
\text{\abbr} w/o Handling
& 21.4 & 1.3h
& 26.9 & 7.9m
& 23.0 & 6.4h
& 30.8 & 1.2h \\
\rowcolor{blue!12}
\textbf{\text{\abbr}}
& \textbf{21.7} & \textbf{1.3h}
& \textbf{27.3} & \textbf{8.0m}
& \textbf{23.2} & \textbf{6.4h}
& \textbf{31.2} & \textbf{1.2h} \\
\bottomrule
\end{tabular}
\end{table*}

\begin{table}[t]
\centering
\caption{Repartitioning overhead across scenes.}
\vspace{-3mm}
\label{tab:reb-repartition-overhead}
\resizebox{\columnwidth}{!}{%
\begin{tabular}{lcccc}
\toprule
 & \makecell{Small City\\Street} & \makecell{Small City\\Aerial} & \makecell{Big City\\Street} & \makecell{Big City\\Aerial} \\
\midrule
Total Training Time   & 1.3h    & 8.0m    & 6.4h     & 1.2h    \\
\# Repartitions       & 7       & 2       & 43       & 11      \\
Avg.\ Repart.\ Time   & 1.5s    & 1.1s    & 3.5s     & 2.1s    \\
Total Repart.\ Overhead & 10.5s & 2.2s    & 150.5s   & 23.1s   \\
Overhead Ratio        & 0.22\%  & 0.46\%  & 0.65\%   & 0.53\%  \\
\bottomrule
\end{tabular}%
}
\end{table}

We monitor load balance during training and trigger repartitioning only when the imbalance becomes significant. Load balance is checked every few thousand iterations via a lightweight all-gather of per-GPU Gaussian counts ($<$100\,ms), and repartitioning is triggered only when the imbalance ratio exceeds 20\% (Table~\ref{tab:reb-partition-sensitivity}).
New boundaries are computed using lightweight local spatial histograms, followed by all-to-all redistribution.
As shown in Table~\ref{tab:reb-repartition-overhead}, the total repartitioning overhead accounts for less than 0.7\% of training time across all scenes, confirming that dynamic repartitioning is effectively free.

\subsection{Experimental Analysis}

\noindent\ul{\textbf{Saturation Stability during Training.}}
\label{reb-E3}
Since opacity evolves during training, pixels marked as saturated may later become informative again, raising the question of whether saturation decisions remain valid over time. We show that frequent re-evaluation is unnecessary in practice: the first two analyses demonstrate that such flip cases are rare and yield only marginal quality gains, while the third provides a low-cost fallback when strict correctness is desired.

\ul{\textit{i) Low flip rate.}} When a GPU is pruned for a given pixel, active GPUs could retain its per-partition \emph{transmittance consumption budget} (stale values from the last visit before pruning) through the normal composition process ($\S$\ref{sec:design-pixel-level}). In subsequent iterations, when the view is rendered again, each active GPU obtains a fresh residual transmittance from its routine global composition and checks whether the residual \emph{budget} exceeds the transmittance threshold $\varepsilon$. If so, we perform a \textit{speculative evaluation} to determine how many pruned GPUs the residual can sustain by sequentially subtracting each pruned GPU's stale consumption budget in depth order until the residual budget falls below $\varepsilon$.
Only those pruned GPUs whose budgets are fully covered are counted as flips. We define the flip rate as the ratio of flipped pixel-GPU pairs to the total number of pruned pairs, which is low across all scenes as summarized in Table \ref{tab:flip-rate}.

\begin{table}
\caption{Flip rate across all scenes.}
\vspace{-3mm}
\label{tab:flip-rate}
\centering
\footnotesize
\begin{tabular}{lcccc}
\toprule
\shortstack{Scene\\~} & \shortstack{Small City\\Street} & \shortstack{Small City\\Aerial} & \shortstack{Big City\\Street} & \shortstack{Big City\\Aerial} \\
\midrule
Flip Rate & 0.42\% & 0.21\% & 0.36\% & 0.18\% \\
\bottomrule
\end{tabular}
\end{table}

\ul{\textit{ii) Marginal performance effects. }} To quantify the practical impact of these flip pairs, we run an upper-bound ablation that \emph{disables saturation pruning}, i.e., always keeps all potentially flipped pairs active. PSNR improves only marginally with less than 0.05dB. The results in \textit{i)} and \textit{ii)} indicate that ignoring flips has a negligible effect in practice and frequent re-evaluation is unnecessary. 

\ul{\textit{iii) Re-evaluation at a low cost.}} For strict correctness, saturation can also be re-evaluated every iteration: the fresh residual transmittance is already produced by routine composition during rendering, and the speculative evaluation is a lightweight local scan on active GPUs. The only additional step is notifying previously pruned GPUs to re-activate when the same view is rendered again, which can be implemented as an asynchronous and non-blocking control message, avoiding any additional synchronization on the critical path.